\documentclass[conference]{IEEEtran}

\usepackage{xspace}
\usepackage{etoolbox}
\makeatletter

\patchcmd{\@makecaption}
  {\scshape}
  {}
  {}
  {}
\makeatletter
\patchcmd{\@makecaption}
  {\\}
  {.\ }
  {}
  {}
\makeatother
\def\tablename{Table}

\usepackage{amsmath,amsfonts}
\usepackage{algorithmic}
\usepackage{graphicx}

\usepackage{textcomp}
\usepackage{todonotes}
\usepackage{booktabs}
\usepackage{tabularx}
\usepackage{multirow}
\usepackage{arydshln}

\usepackage{xurl}
\usepackage{balance}
\usepackage{float}
\usepackage{stfloats}
\usepackage{tikz}
\usepackage{xcolor}
\usepackage{bbding} 
\usepackage{threeparttable} 
\usepackage{makecell}

\usepackage{array}
\newcolumntype{L}[1]{>{\raggedright\let\newline\\\arraybackslash\hspace{0pt}}m{#1}}
\newcolumntype{C}[1]{>{\centering\let\newline\\\arraybackslash\hspace{0pt}}m{#1}}
\newcolumntype{R}[1]{>{\raggedleft\let\newline\\\arraybackslash\hspace{0pt}}m{#1}}

\DeclareRobustCommand*{\circled}[1]{\lower.7ex\hbox{\tikz\draw[fill=black] (0pt, 0pt) circle (.43em) node {\makebox[0.7em][c]{\small\textcolor{white}{#1}}};}}

\def\BibTeX{{\rm B\kern-.05em{\sc i\kern-.025em b}\kern-.08em
    T\kern-.1667em\lower.7ex\hbox{E}\kern-.125emX}}

\definecolor{brickred}{rgb}{0.8, 0.25, 0.33}

\providecommand\wenhao[1]{\textcolor{blue}{\{\textbf{wenhao:} {\em#1}\}}}

\newcommand{\sysname}{NestedSGX\xspace}

\newenvironment{packeditemize}{
\begin{list}{$\bullet$}{
\setlength{\labelwidth}{2pt} 
\setlength{\itemsep}{0pt}
\setlength{\leftmargin}{\labelwidth}
\addtolength{\leftmargin}{\labelsep}
\setlength{\parindent}{0pt}
\setlength{\listparindent}{\parindent}
\setlength{\parsep}{1pt} 
\setlength{\topsep}{1pt}}}{\end{list}}

\DeclareRobustCommand*{\authorrefmark}[1]{\raisebox{0pt}[0pt][0pt]{\textsuperscript{\footnotesize\ensuremath{\ifcase#1\or *\or \dagger\or \ddagger\or%
    \mathsection\or \mathparagraph\or \|\or **\or \dagger\dagger%
    \or \ddagger\ddagger \else\textsuperscript{\expandafter\romannumeral#1}\fi}}}}
\usepackage{bbding} 

\definecolor{indianred}{rgb}{0.8, 0.36, 0.36}
\usepackage{hyperref}
\hypersetup{
    colorlinks = true,
    allcolors = magenta,
}

\usepackage{caption}
\usepackage{subcaption}

\newcommand{\para}[1]{\vspace{3pt}\noindent\textit{\textbf{{#1. }}}}
\providecommand\ignore[1]{{}}

\usepackage{cite}
\begin{document}

\title{
The Road to Trust: Building Enclaves within Confidential VMs
}

\author{
{\rm  Wenhao Wang\authorrefmark{1}\authorrefmark{4}, Linke Song\authorrefmark{1}\authorrefmark{4}, Benshan Mei\authorrefmark{1}\authorrefmark{4}, Shuang Liu\authorrefmark{2}, Shijun Zhao\authorrefmark{1},} \\ {\rm  Shoumeng Yan\authorrefmark{2}\textsuperscript{\Envelope}, XiaoFeng Wang\authorrefmark{3}, Dan Meng\authorrefmark{1}, Rui Hou\authorrefmark{1}\authorrefmark{4}\textsuperscript{\Envelope}}\\ \\
\authorrefmark{1}Key Laboratory of Cyberspace Security Defense, Institute of Information Engineering, CAS \\
\authorrefmark{2}Ant Group \\
\authorrefmark{3}Indiana University Bloomington \\
\authorrefmark{4}School of Cyber Security, University of Chinese Academy of Sciences \\}

\ignore{
\author{}
\author{Wenhao Wang}
\affiliation{%
   \institution{Institute of Information Engineering, Chinese Academy of Sciences}
   \country{}
}
\email{wangwenhao@iie.ac.cn}

\author{Linke Song}
\affiliation{%
   \institution{Institute of Information Engineering, Chinese Academy of Sciences}
   \country{}
}
\email{songlinke@iie.ac.cn}

\author{Benshan Mei}
\affiliation{%
   \institution{Institute of Information Engineering, Chinese Academy of Sciences}
   \country{}
}
\email{meibenshan@iie.ac.cn}

\author{Shuang Liu}
\affiliation{%
   \institution{Ant Group}
   \country{}
}
\email{ls123674@antgroup.com}

\author{Shijun Zhao}
\affiliation{%
   \institution{Institute of Information Engineering, Chinese Academy of Sciences}
   \country{}
}
\email{zhaoshijun@iie.ac.cn}

\author{Shoumeng Yan}
\affiliation{%
   \institution{Ant Group}
   \country{}
}
\email{shoumeng.ysm@antgroup.com}

\author{XiaoFeng Wang}
\affiliation{%
   \institution{Indiana University Bloomington}
   \country{}
}
\email{xw7@indiana.edu}

\author{Dan Meng}
\affiliation{%
   \institution{Institute of Information Engineering, Chinese Academy of Sciences}
   \country{}
}
\email{mengdan@iie.ac.cn}

\author{Rui Hou}
\affiliation{%
   \institution{Institute of Information Engineering, Chinese Academy of Sciences}
   \country{}
}
\email{hourui@iie.ac.cn}
\renewcommand{\shortauthors}{Wenhao Wang et al.}
}

\IEEEoverridecommandlockouts
\makeatletter\def\@IEEEpubidpullup{6.5\baselineskip}\makeatother
\IEEEpubid{\parbox{\columnwidth}{
    Network and Distributed System Security (NDSS) Symposium 2025\\
    24 February - 28 February 2025, San Diego, CA, USA\\
    ISBN x-xxxxxx-xx-x\\
    https://dx.doi.org/xx.xxxxx/ndss.2025.xxxxx\\
    www.ndss-symposium.org
}
\hspace{\columnsep}\makebox[\columnwidth]{}}

\maketitle 

\begin{abstract}
Integrity is critical for maintaining system security, as it ensures that only genuine software is loaded onto a machine. Although confidential virtual machines (CVMs) function within isolated environments separate from the host, it is important to recognize that users still encounter challenges in maintaining control over the integrity of the code running within the trusted execution environments (TEEs). The presence of a sophisticated operating system (OS) raises the possibility of dynamically creating and executing any code, making user applications within TEEs vulnerable to interference or tampering if the guest OS is compromised.
To address this issue, this paper introduces \sysname, a framework which leverages virtual
machine privilege level (VMPL), a recent hardware feature available on AMD SEV-SNP to enable the creation of hardware enclaves within the guest VM. Similar to Intel SGX, \sysname considers the guest OS untrusted for loading potentially malicious code. It ensures that only trusted and measured code executed within the enclave can be remotely attested. To seamlessly protect existing applications, \sysname aims for compatibility with Intel SGX by simulating SGX leaf functions. We have also ported the SGX SDK and the Occlum library OS to \sysname, enabling the use of existing SGX toolchains and applications in the system. Performance evaluations show that context switches in \sysname take about 32,000 -- 34,000 cycles, approximately $1.9\times$ -- $2.1\times$ higher than that of Intel SGX. \sysname incurs minimal overhead in most real-world applications, with an average overhead below 2\% for computation and memory intensive workloads and below 15.68\% for I/O intensive workloads.
\end{abstract}


\section{Introduction}
\label{sec:intro}

Confidential computing enhances cloud security by allowing tenants to control the trusted components of their workloads. It minimizes the need for trust in hardware, software, and services, while providing strong protection against attacks from other tenants and the cloud provider. This empowers tenants to develop and deploy confidential applications for their most sensitive data.

At the core of the confidential computing stack is the trusted execution environment (TEE), which isolates the code and data of a confidential workload from other code running on the system, even at the highest privilege levels. In particular, Intel SGX~\cite{sgxoverview} provides protection for the trusted components of an application, known as enclaves, allowing only measured code to run within the TEEs\footnote{Intel SGXv2~\cite{mckeen2016edmm} enables dynamic changes to the page attribute and dynamic loading of code, but with strict controls in place.}. On the other hand, VM-based TEEs, such as AMD SEV~\cite{sev2020strengthening}, Intel TDX~\cite{tdx2020}, and ARM CCA~\cite{armcca} offer a fully backward-compatible confidential virtual machine (CVM), enabling the execution of existing applications without modifications. However, placing entire VMs within TEEs is generally considered less desirable: a VM image is far more than just a kernel and an application -- it includes a large number of system services. It remains uncertain whether it is more secure than running the software on premises or on existing cloud infrastructure. 
Therefore, an interesting question still remains open for VM-based TEEs: 

\begin{center}
\emph{How can users attest and establish trust in applications running within a CVM, given the dynamic nature of building, loading, modification, and execution of arbitrary code within the CVM, if the guest OS might be compromised?}
\end{center}

A seminal work called vSGX~\cite{zhao2022vsgx} was developed to virtualize Intel SGX on AMD SEV, ensuring full binary compatibility and enabling the execution of unmodified SGX programs in CVMs. vSGX adopts a two-VM approach where one VM handles the untrusted application while the other hosts the enclave. However, this emulation of SGX leaf instructions incurs significant overhead, particularly for control flow transfers across the boundary of the two VMs (e.g., EENTER and EEXIT).
For instance, an empty ECall operation on vSGX takes approximately 1.5ms, which is around $160\times$ slower than on native Intel SGX. 
This slow context switch severely limits the service throughput achievable within the enclave. In practical scenarios, this limitation becomes a critical performance bottleneck. To illustrate, vSGX is $6\times$ slower than the baseline on the cURL benchmark, and launching a 256 MB size vSGX enclave takes about 5 minutes.


\begin{figure}
    \centering
    \includegraphics[width=0.9\linewidth]{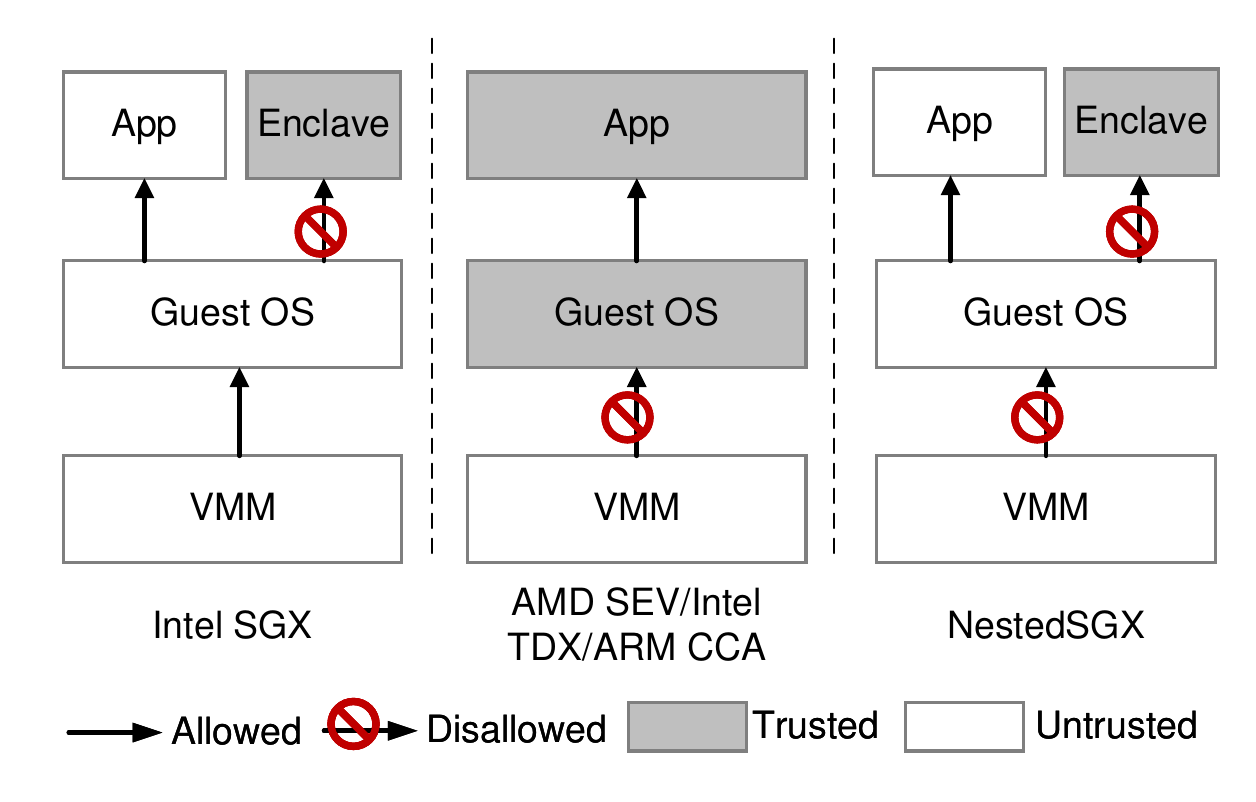}
    \caption{Comparison between \sysname and other TEEs. \sysname offers a layered protection mechanism against both the host VMM and the guest OS within the CVM.}
    \label{fig:comparison}
\end{figure}


\para{Design}
We take a further step in this direction
by introducing our system, called \sysname. \sysname follows a defense-in-depth approach to attesting applications (as depicted in Fig.~\ref{fig:comparison}), where both the enclaves and the guest OS run \textit{within the same CVM}.
Specifically, the trusted portion of an application (i.e., the enclave) is loaded and executed within an isolated environment, separated from the feature-rich OS, similar to the enclave abstraction of Intel SGX. During the loading process, the trusted portion is measured and can be attested to a remote user. The core root of trust for measurement (CRTM) and root of trust for reporting (RTR) rely on a small segment of \textit{trusted code}, known as the security monitor. This security monitor enforces memory isolation
within the CVM,
and its integrity is also measured and attested with the assistance of hardware-based TEE.
This approach empowers us to dynamically measure and attest the application code, enabling us to bypass the complexities associated with the guest OS.
To facilitate the protection of existing applications, \sysname prioritizes compatibility with Intel SGX. This involves simulating SGX leaf functions with a trusted SGX emulation layer to manage the entire life cycle of enclaves, including creation, memory management, teardown, and context switches. 

In more details, our approach leverages the privilege separation offered by AMD SEV-SNP, particularly the virtual machine privilege level (VMPL), to establish isolated guest physical space exclusively occupied by the security monitor and the enclaves. With VMPL, the vCPUs within the CVM are configured to operate at different privilege levels. Each privilege level is associated with specific access permissions to the guest physical memory pages. This capability enables us to assign distinct privilege levels to different software components running within the CVM. Particularly, the security monitor, the SGX emulation layer and the enclave operate at the higher VMPL (e.g., VMPL0), in the kernel mode and user mode respectively. On the other hand, the guest OS and the untrusted part of the application (App) are required to run in the kernel mode and user mode at a lower VMPL (e.g., VMPL1). This design ensures that the guest OS does not have access to the memory allocated for VMPL0.




We implemented \sysname on top of AMD's Linux Secure VM Service Module (SVSM) framework~\cite{svsmgithub}, without necessitating any modifications of the host software, such as KVM. 
Furthermore, we have successfully ported the Intel SGX SDK~\cite{sgxsdk}, the Rust SGX SDK~\cite{wang2019towards} as well as the Occlum library OS~\cite{shen2020occlum} to the \sysname framework, allowing for smooth integration of existing SGX toolchains and applications. 
In addition, we have implemented the HotCalls~\cite{weisse2017regaining} optimization in Occlum, which allows specific OCALLs to be handled asynchronously within the enclave without exiting it. This optimization has led to improved performance for applications running on Occlum without requiring any modifications to their source code. Unlike vSGX, we stress that
\textit{full binary compatibility with SGX is a non-goal of our paper}.

We conducted a series of benchmarking tests on commercial AMD SEV-SNP hardware and observed that the cost of world switches is about $1.9\times$ -- $2.1\times$ higher than Intel SGX, which is nearly \textit{two orders of magnitude faster than vSGX}. Moreover, \sysname introduces negligible overhead in most real-world applications, with less than 2\% overhead for memory and computation-intensive tasks and less than 15.68\% overhead for I/O-intensive tasks.
Generally, \sysname exhibits moderate costs that are comparable to other enclave systems, such as Intel SGX. Consequently, we view it as a promising solution for addressing the critical challenge of performing secure computations within CVMs.







\para{Contributions} The contributions are summarized as follows.

\begin{packeditemize}
    \item \textit{Design}. We present the design of \sysname, incorporating the VMPL feature within AMD SEV-SNP. \sysname addresses the challenge of establishing trust for applications while minimizing reliance on the potentially compromised guest OS.
    
    \item \textit{Compatibility with Intel SGX}. \sysname offers compatibility with Intel SGX, allowing for similar management of enclaves. It significantly reduces the effort required to port existing applications onto the \sysname framework.
    \item \textit{Implementations and evaluations}. We implemented \sysname on commercial hardware. The evaluation demonstrates the overhead incurred by \sysname is comparable to that of Intel SGX, affirming its efficiency and viability.
\end{packeditemize}

\para{Roadmap}
The rest of the paper is organized as follows. We provide necessary background information about Intel SGX and SEV-SNP, along with the threat model in Sec.~\ref{sec:background}. We present the detailed design of \sysname in Sec.~\ref{sec:design}, followed by implementation details in Sec.~\ref{sec:impl}. The security analysis and performance evaluation of \sysname are provided in Sec.~\ref{sec:security} and Sec.~\ref{sec:eval}, respectively. We discuss possible extensions and related works of \sysname in Sec.~\ref{sec:discuss} and Sec.~\ref{sec:related}. Sec.~\ref{sec:conclusion} concludes the paper.
\section{Background}
\label{sec:background}

\subsection{Intel SGX}

Intel Software Guard Extensions (SGX) is a hardware-based security technology designed to provide a trusted execution environment (TEE) called an enclave for secure processing of sensitive computations. Enclaves are isolated regions within the application's virtual address space, sharing the same page table with the untrusted part of the application.
The enclave page cache (EPC) is a dedicated pool of physical memory reserved for enclaves. All enclave pages are encrypted by the hardware, safeguarding against unauthorized access even if an adversary gains physical access to the memory.
To mitigate memory mapping attacks by manipulating the page table, the hardware manages the enclave page cache metadata (EPCM). This metadata stores essential information for each EPC page, such as allocation state, owner, and access permissions. During page table walk, the EPCM is consulted to ensure that only authenticated operations are permitted.

\ignore{
\begin{figure}
    \centering
    \includegraphics[width=1\linewidth]{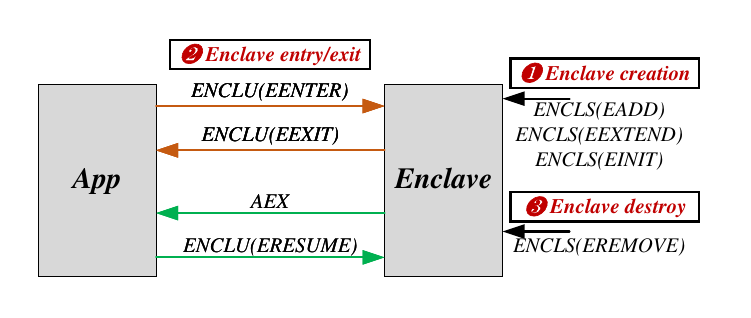}
    \caption{The enclave's life cycle in Intel SGX.}
    \label{fig:sgx_life_cycle}
\end{figure}
}

The SGX hardware enables user space applications to set aside private memory regions of code and data. The allocation of the private memory is managed through a set of privileged (ring-0) ENCLS leaf functions, while a set of unprivileged (ring-3) ENCLU leaf functions allow applications to enter and execute within these regions. 
Additionally, the hardware records meta-information for the enclave and its thread to effectively manage the enclave's life cycle.
In the event of interrupts or exceptions, the hardware triggers an Asynchronous Enclave Exits (AEX) event. During an AEX, the enclave's context is saved and the application context is restored. Upon handling the interrupt or exception, the enclave context is restored when the execution resumes within the enclave.


\para{SGX attestation}
During enclave creation, the hardware measures the integrity of the enclave code and data. This allows for remote attestation, enabling external entities to verify the trustworthiness of the enclave. In most cases, \textit{every component of enclave code can be measured and attested}. The root of trust for reporting resides in the attestation key, which is exclusively accessible by the Intel signed quote enclave (QE).

\subsection{SEV-SNP and VMPL}
\label{subsec:snp}

AMD provides confidential computing capabilities through secure encrypted virtualization (SEV), a technology that builds upon virtualization. SEV utilizes a dedicated security processor -- AMD Security Processor (AMD-SP) -- with independent firmware, distinct from the primary x86 cores.
SEV-ES, an enhancement to SEV, goes beyond encrypting guest memory and also encrypts the guest CPU registers. This ensures that in-flight values are protected from leakage and prevents malicious manipulation of registers by a compromised hypervisor.


SEV-SNP, introduced in 2020~\cite{sev2020strengthening}, is the third generation of SEV and provides enhanced security against malicious manipulation of page mappings by the host. One of its key features is the Reverse Map Table (RMP), a structure located in secure memory that maps system physical addresses (sPAs) to guest physical addresses (gPAs). The RMP serves as a metadata table managed by the AMD-SP and plays a critical role in tracking the ownership of each system physical page to prevent the host from writing to encrypted guest pages. It establishes a global one-to-one mapping between sPAs and gPAs, ensuring that a page cannot be simultaneously mapped into multiple guests or multiple times within a single guest.


Virtual Machine Privilege Level (VMPL) is an optional feature within the SEV-SNP architecture that allows a CVM to divide its address space into four distinct levels. These levels serve as hardware-isolated abstraction layers for the CVM, with VMPL0 representing the highest privilege level and VMPL3 representing the lowest. Each hardware context, known as a Virtual Machine Save Area (VMSA), is associated with a specific VMPL. Moreover, different memory pages assigned to a guest can have varying permissions based on the VMPL.
VMPLs are utilized for additional page permission checks and are independent of other x86 security features.

The RMP maintains records of the read, write, and execution permissions for each guest physical page across different VMPLs. During the translation of a virtual address to a physical address by the CPU or IOMMU, an RMP check is typically conducted to determine the relevant permissions and ownership of each physical memory page.
The guest has the ability to modify VMPL permissions using the \texttt{RMPADJUST} instruction. This instruction allows a higher privilege VMPL to adjust the permissions of a lower privilege VMPL. It grants the guest the flexibility to modify and manage the permissions of different VMPLs based on its specific requirements.
In addition to SEV-SNP, AMD provides the Secure Virtual Machine Service Module (SVSM) framework~\cite{svsm2022spec}, which is a small piece of code running at the highest privilege level (VMPL0) to provide security services (such as virtual TPM) to the rest of the guest.


\para{Switching VMPLs}
VMPL switching requests are triggered by the CVM through the \texttt{VMGEXIT} instruction and captured by the hypervisor. The hypervisor switches the execution to the targeted VMPL through the \texttt{VMRUN} instruction, with the VMSA associated with the targeted VMPL as the parameter. There are two methods for the CVM to communicate the desired VMPL to the hypervisor, either by using the MSR protocol with a value of the requested VMPL in the GHCB MSR (MSR 0xc0010130), or by setting the shared GHCB with the exit code \path{GHCB_NAE_RUN_VMPL} (0x80000018) and the desired VMPL as the exit information.

\ignore{
\begin{figure}
    \centering
    \includegraphics[width=0.9\linewidth]{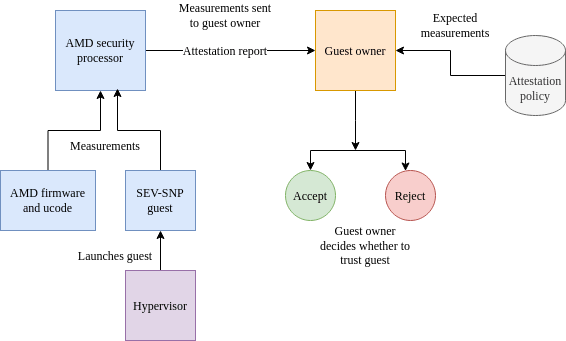}
    \caption{SEV-SNP attestation flow.}
    \label{fig:snp-attestation}
\end{figure}
}

\para{SEV-SNP attestation}
The SEV-SNP CVM is initialized from an unencrypted initial image. This image contains the boot code necessary for the CVM but does not include any confidential information. During the launch process, the hypervisor requests the AMD-SP to install this initial set of pages in the CVM. The AMD-SP cryptographically measures the content of these pages, along with the associated metadata, to ensure an accurate measurement of the initial guest memory layout.
Following that, the boot code triggers the bootstrapping of the OS image, which is also a part of the measured initial image. The attestation report, signed by the AMD-SP firmware, allows a third party, such as the guest owner, to verify the state of the CVM running on an authentic AMD platform.



\subsection{Threat model}
\label{subsec:threat}

\sysname is built on top of SEV-SNP and follows a similar threat model, which considers the host software as a potential source of harm or compromise. We place our trust in the underlying SEV-SNP hardware to protect the CVM from direct observation or tampering by the host software or other CVMs. Specifically, we do not consider the recent attacks on SEV-SNP, such as the CacheWrap attack~\cite{Zhang2024CacheWarp}, the WeSee attack~\cite{schluter2024wesee} and the Heckler attack~\cite{schluter2024heckler}. 

Since the user can attest the load-time correctness of the CVM with remote attestation, we assume that the attacker initially only controls all software and hardware external to the CVM. However, considering the substantial code base of the guest OS within the CVM, it is not considered entirely trustworthy. Therefore, the primary goal of \sysname is to strongly maintain the integrity and confidentiality of the enclave, effectively protecting it from potential security compromises originating from both the host software and the untrusted guest OS. These entities may collaborate in their attempts to breach the enclave's security.

Furthermore, this paper does not 
cover 
the security risks associated with unsafe code, such as {memory safety vulnerabilities within the enclave~\cite{biondo2018guard}, enclave malwares~\cite{toffalini2021snakegx,schwarz2017malware},} or attacks originating from the interface between the enclave and the application (e.g., COIN~\cite{khandaker2020coin} or IAGO~\cite{checkoway2013iago} attacks). However, it is important to highlight that \sysname ensures that the enclave cannot compromise the operations or integrity of the guest OS, even in the presence of vulnerabilities.
The security monitor is responsible for enforcing memory isolation between mutually distrustful components. The SGX emulation layer serves as the core root of trust for measurement (CRTM). To establish a foundation of trust, it is assumed that the security monitor and the SGX emulation layer are trusted. 

This paper does not cover side channel attacks, including traditional cache side channel attacks~\cite{liu2015last}, page table based attacks~\cite{xu2015controlled,wang2017leaky,van2017telling} or transient execution attacks~\cite{kocher2020spectre,van2018foreshadow,van2019ridl}. We consider protection against these types of attacks as orthogonal to our design. Furthermore, since the host and the guest OS are free to not schedule the enclave for execution, Denial-of-Service (DoS) attacks are also out of scope.
\section{Design}
\label{sec:design}

\subsection{Overview}
\label{subsec:overview}

\begin{figure}
    \centering
    \includegraphics[width=\linewidth]{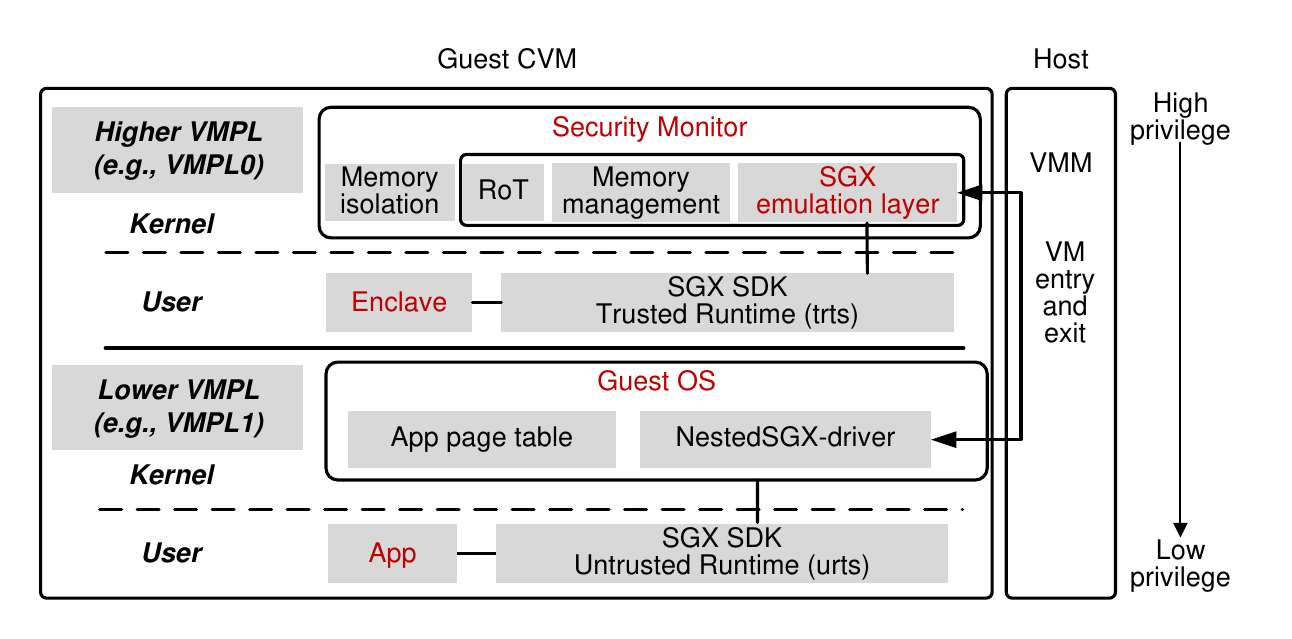}
    \caption{\sysname overview. The security monitor, the SGX emulation layer and the enclave operate in the kernel and user mode at VMPL0, while the guest OS and App operate in the kernel and user mode at VMPL1. The transitions between VMPL0 and VMPL1 occur via the untrusted host.}
    \label{fig:overview}
\end{figure}

As shown in Fig.~\ref{fig:overview}, \sysname encompasses the following components within the CVM. 
The application is divided into the trusted part (\textit{\textbf{enclave}}) and the untrusted part (\textit{\textbf{App}}). 
The \textit{\textbf{security monitor}} is responsible for enforcing security isolation within \sysname.
The \textit{\textbf{SGX emulation layer}} emulates SGX instructions, and serves as the root of trust (RoT) for measurement and reporting of the enclave.
The \textit{\textbf{guest OS}}
offers essential features to applications, such as file systems, networking, device drivers, and language runtime.

To enforce memory isolation, the security monitor runs at the highest privileged VMPL (i.e., VMPL0). On the other hand,
we cannot place the SGX emulation layer in a VMPL lower than that of the guest OS (e.g., guest OS in VMPL1 and SGX emulation layer in VMPL2). Doing so would grant the guest OS the ability to generate attestation reports on behalf of the SGX emulation layer, potentially compromising the trust chain (please refer to Sec.~\ref{subsec:attestation} for details).
Within our threat model, we consider the SGX emulation layer to be trusted. 
For the sake of brevity, throughout the remainder of the paper, we regard \textit{the SGX emulation layer as an integral component of the security monitor}, both running at VMPL0.
while the guest OS and App operate at a lower privileged VMPL (i.e., VMPL1). 
Within this design, we also place the enclaves within the same VMPL as the SGX emulation layer, in the user mode and kernel mode respectively, since if they were placed at different VMPLs, a context switch between the enclave and App would necessitate 2 costly VMPL switches.
In practice, however, if running both the SGX emulation layer and the enclaves in VMPL0 raises security concerns, we can adapt NestedSGX slightly as follows: the security monitor still runs in VMPL0; the SGX emulation layer and the enclaves run in the kernel and user mode at VMPL1, and the guest OS and the App run in the lowest VMPL (e.g., VMPL3).


To let the users maintain control over the code running inside the CVM and bootstrap trust to the enclave within a feature-rich guest OS, \sysname is designed to fulfill the following requirements. 

\textit{Firstly}, \sysname ensures that any transitions between the trusted and untrusted components are mediated by the security monitor. This guarantees that sensitive information is properly sanitized and safeguarded. To leverage existing SGX toolchains, \sysname employs the SGX programming model to manage the enclave's life cycle.
The details for secure enclave state transitions will be presented in Sec.~\ref{subsec:transitions}.

\textit{Secondly}, the system establishes isolated memory regions exclusively used by the security monitor, the enclaves and the guest OS respectively. This ensures that the enclave and the security monitor are secure even in scenarios where the guest OS operates in kernel mode. Specifically, the entire gPA space is divided to two parts: a continuous and fixed region exclusively used by VMPL0, and the remaining portion used by VMPL1. 
Additionally, our design guarantees that the enclave cannot access the memory of the guest OS, despite running at a higher VMPL.
The details of the memory isolation scheme will be presented in Sec.~\ref{subsec:isolation}.

\textit{Lastly}, in \sysname, only authenticated code and data can be loaded and processed within the enclave. 
To achieve this, \sysname enforces a mechanism where the enclave can only be loaded by the trusted security monitor, which also performs the measurement of the enclave's integrity.
While the guest OS retains the freedom to load and execute any code within its own address space, these code segments are not affirmed by the security monitor and therefore will not be attested.
The details of enclave measurement and attestation will be presented in Sec.~\ref{subsec:attestation}.


\begin{table*}
    \centering
    \caption{Comparison with competing approaches.}
    \label{tab:comparison}
    \begin{tabular}{c C{2.0cm} C{2.0cm}  C{2.0cm}  C{2.0cm} C{2.0cm} C{2.0cm}}
    \toprule
         & \textbf{Unlimited enclaves} & \textbf{Multi-threading} & \textbf{Exception handling} & \textbf{SGX ecosystem} & \textbf{No changes to hypervisor} & \textbf{Low-overhead context switches}  \\ \midrule
        vSGX~\cite{zhao2022vsgx} & \XSolidBrush & \Checkmark & \Checkmark  & \Checkmark  & \XSolidBrush & \XSolidBrush \\ \midrule
        Veil~\cite{ahmad2023veil} & \Checkmark  & \XSolidBrush & \XSolidBrush & \XSolidBrush & \XSolidBrush & \Checkmark  \\ \midrule
        \sysname & \Checkmark  & \Checkmark  & \Checkmark  & \Checkmark & \Checkmark & \Checkmark \\ \bottomrule
    \end{tabular}
\end{table*}

\para{Comparison with competing approaches (Table~\ref{tab:comparison})}
vSGX~\cite{zhao2022vsgx} effectively virtualizes Intel SGX on AMD SEV, ensuring complete binary compatibility and the execution of unmodified SGX programs. vSGX employs two separated CVMs to accommodate the untrusted application and the enclave. In contrast, \sysname adopts a more intuitive design that utilizes different VMPLs for memory isolation. The \sysname design offers the following benefits.

Firstly, vSGX cannot support directly sharing encrypted memory between the enclave VM (EVM) and application VM (AVM). To provide confidentiality, integrity and replay protection, a dedicated communication protocol and encrypted channel are essential. Within \sysname, the security monitor can access the entire guest physical address space, and communicate with the guest OS through shared memory. This approach circumvents the overhead associated with memory encryption and movement, and prevents the hypervisor from observing traffic patterns. As a comparison, an empty ECALL costs 1.5 ms on vSGX, and only 12 us on \sysname, which is two orders of magnitude faster than vSGX.

Secondly, with vSGX, only one enclave can be hosted within the EVM, which restricts the maximum number of concurrently-running enclaves on the hardware platform. This limitation arises because SEV associates ASIDs with VMs' memory encryption keys, and the ASID bit is limited. In contrast, \sysname has the ability to support an unlimited number of enclaves.


Concurrent to our work, 
Ahmad et al. proposed Veil~\cite{ahmad2023veil} as a service framework running at VMPL0. 
As one use case, Veil offers protection to the entire applications within a separated VMPL, facilitating redirection of system calls and interrupts. {However, its design has some limitations: (1) Exception handling requires enclave context information and is not supported; (2) On context switches (e.g., syscall and interrupt), VMPL switches to the guest OS are triggered by the enclave, and don't follow the standard GHCB protocols, necessitating changes to the hypervisor; (3) Veil does not support SGX ecosystems.}
{Addressing these limitations requires an SGX emulation layer to emulate SGX instructions and fulfill certain OS functionalities, such as mediating all enclave-App context switches and forwarding exceptions to the guest OS, within the user-land enclave’s VMPL. Designing and implementing this layer is challenging because its state -- including general-purpose and segment registers, stack pointers -- changes during context switches when handling various requests (e.g., forwarding exceptions or emulating SGX instructions), so when the system re-enters the enclave again, its state cannot be fully restored due to the change of the state in the emulation layer that handles the context switch. As a result, the handling of subsequent requests can lead to failures, such as faults and stack exhaustion.}

{Note that this challenge cannot be resolved by vSGX or Veil. vSGX runs the App and enclave in different VMs and can simply pause or activate one of them for context switching, but also incurs significant overheads for cross-VM communications during the switch. Context switches in Veil rely on hardware, which automatically saves and restores the contexts of the enclave and the guest OS within their respective VMSAs. But this mechanism cannot handle the switches triggered by exceptions.}

{Our solution introduces a unique design that models the states of the SGX emulation layer using a finite state machine (FSM). After each context switch, the layer restores its original state under the FSM's guidance. We achieve this by pairing asynchronous and synchronous enclave entry and exit requests. We save the state of the SGX emulation layer before the entry request (EENTER/AEX) and restore it after processing the exit request, through a carefully designed restoration strategy based upon the FSM (e.g., restoring gs before rsp to balance the stack). In this way, the enclave state can be guaranteed to be correct no matter how complicated the request sequence could become (e.g., arbitrary exceptions between EENTER and EEXIT). This design, along with our thorough implementation of exception handling and multithreading, differentiates \sysname from both Veil and vSGX.}

\subsection{Enclave life cycle management}
\label{subsec:transitions}

In \sysname, the life cycle of an enclave closely resembles that of SGX. Initially, the application invokes the ECREATE leaf function to create the enclave, which initializes the SGX Enclave Control Structure (SECS) page. Subsequently, EPC pages necessary for the enclave, such as code sections, data sections, and Thread Control Structure (TCS) pages, are added using EADD and measured using EEXTEND leaf functions.
Once all the required EPC pages are added, the enclave is initialized with EINIT, and the final measurement is performed. Upon completion of the enclave's execution, the EPC pages are reclaimed using the EREMOVE leaf function, and the enclave is destroyed.

\sysname achieves compatibility with SGX by transitioning to the security monitor on these leaf functions. Since VMPL switches can only be executed in the privileged mode, \sysname provides the \sysname-driver within the guest OS, which accepts the request from the App and facilitates the VMPL switch. During this process, the \sysname-driver receives the type of SGX leaf functions, while the parameters are saved on the App's stack. The \sysname-driver then switches the execution to VMPL0 and transfers control to the security monitor, where the security monitor emulates the instructions strictly adhering to the SGX specifications.



\begin{figure*}
    \centering   
    \includegraphics[width=\linewidth]{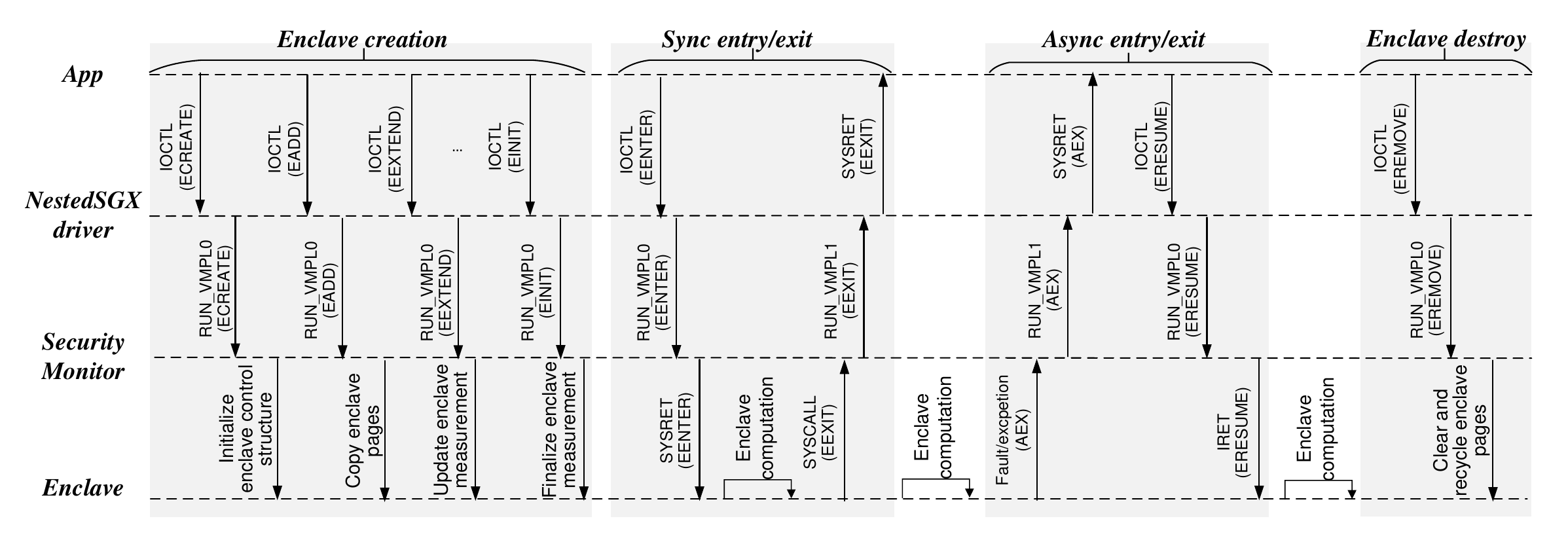}
    \caption{The management of enclave life cycles, e.g., handling synchronous and asynchronous enclave entry and exit events.}
    \label{fig:lifecycle}
\end{figure*}

\para{Synchronous enclave entry and exit}
Once an enclave is initialized, the App can enter the enclave using the EENTER leaf function and jump to the enclave's code for execution. 
As shown in Fig.~\ref{fig:lifecycle}, during the emulation of EENTER, the security monitor switches to the enclave's page table and enters the enclave entry point specified by the TCS using the \texttt{sysret} instruction. The enclave can then return to the App using the EEXIT leaf function, which is also emulated by the security monitor. Specifically, the enclave returns to the security monitor using the \texttt{syscall} instruction. Subsequently, the security monitor switches the execution to the \sysname-driver running at VMPL1. The \sysname-driver restores the App's page table and returns control to the App using the \texttt{sysret} instruction.

\para{Asynchronous enclave entry and exit}
During enclave execution, faults and exceptions may occur. According to the SGX model, these faults and exceptions are supposed to be handled by the guest OS first.
In \sysname, when a fault or exception occurs within the enclave, it traps to the fault and exception handler within the security monitor (Fig.~\ref{fig:lifecycle}). The security monitor then emulates the Asynchronous Exit (AEX) event by saving the context of the enclave in the state saving area (SSA). The security monitor switches the vCPU to the \sysname-driver, which operates at VMPL1. The \sysname-driver fills the instruction pointer (RIP) to the trampoline area (AEP)
and invokes the handler of the guest OS. After the OS handles the fault or exception, it uses the \texttt{iret} instruction to return to the AEP. The AEP, in turn, resumes the enclave execution with the ERESUME leaf function. 
The emulation of the ERESUME leaf function is similar to that of EENTER, except that the enclave's execution context is restored from the SSA.
As a result, \sysname is able to handle exceptions that occur during enclave execution. For example, \sysname supports the emulation of the CPUID instruction, enabling the execution of unmodified applications.
\footnote{In comparison, vSGX~\cite{zhao2022vsgx} requires modifying the application code in order to bypass the CPUID check.}

If the untrusted \sysname-driver chooses not to fill RIP with AEP, this decision does not introduce any new security issues. This is because the control flow outside the enclave is already vulnerable to manipulation within the SGX model and is not inherently trusted.
In \sysname, faults and exceptions are firstly trapped to the security monitor, enabling easy detection of suspicious behaviors such as abnormal interrupts caused by page fault-based attacks. 
Conceptually, \sysname can also support the recent AEX-notify feature~\cite{constable2023aex},
although the implementation is planned for future work.


\subsection{Memory isolation}
\label{subsec:isolation}

\begin{figure}
    \centering
    \includegraphics[width=\linewidth]{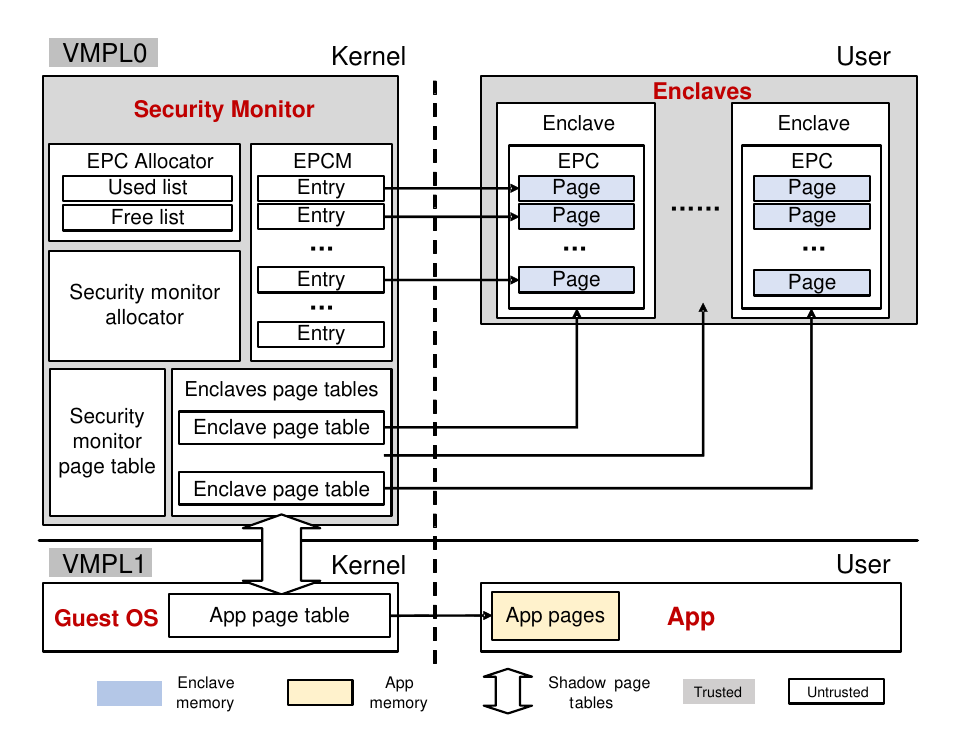}
    \caption{The guest physical address space is divided into 3 parts: the secure memory used by the security monitor, the EPC memory, and the normal memory.}
    \label{fig:memory}
\end{figure}

The entire guest physical address space of the CVM is divided into 3 parts (Fig.~\ref{fig:memory}): the secure memory used by the security monitor, the secure memory used by the enclaves (i.e., EPC), and the normal memory used by the guest OS and untrusted applications. The access to the secure memory is restricted solely to VMPL0, whereas the normal memory can be accessed by both VMPL0 and VMPL1. Notably, the SGX control structures are stored in the memory of the security monitor, while the enclave's code and data pages are stored in the memory reserved for the EPC.

\para{Booting \sysname}
Upon receiving a request to launch a CVM, the platform proceeds to load the VM image and cryptographically measures its contents. During the initialization process, the entire gPA is divided into two distinct parts. By configuring the RMP attributes, we establish a dedicated region of guest memory exclusively reserved for VMPL0, while the remaining gPA regions are reserved for VMPL1. The memory for VMPL0 is further divided into two parts, one for the security monitor and one for the enclaves.

Once the guest image is loaded, the hypervisor places the vCPU in VMPL0 mode and transfers control to the security monitor, which is positioned at the predetermined gPA location. The security monitor takes responsibility of initializing the guest CPU, memory, and setting up a page table for execution. It subsequently hands over control to the BIOS code to initiate the booting process of the guest OS. The security monitor advertises its presence and the memory range reserved for VMPL0. This prevents the guest OS from attempting to utilize any memory within that range. Any such attempts would be detected and blocked by the VMPL permission check, leading to RMP faults. In this manner, \sysname ensures that the guest OS cannot access or interfere with the memory of the security monitor or the enclaves.

\para{EPC memory management}
In Intel SGX, the enclave and application share the same page table, and the hardware prevents memory mapping attacks by performing additional security checks during a page table walk, using the EPCM to ensure that the memory mappings are correct. 
Without hardware support, it is necessary to prevent the guest OS from manipulating the mappings of the enclave. \sysname employs a shadow page table scheme: the application's page table is managed by the guest OS, while the enclave’s page table is managed by the security monitor. In situations where the application and enclave need to share memory, such as for parameter passing, \sysname maps the shared memory to both page tables with the same gVA-to-gPA mapping. The mapping of the parameter buffer remains fixed throughout the entire lifetime of the enclave. Consequently, we do not need to maintain the synchronization of the two page tables.


In \sysname, the security monitor is responsible for managing the EPC memory. It maintains two lists: a free list and a used list, containing all the EPC pages. Similarly, an EPCM data structure is utilized to track the state of each EPC page. During enclave initialization, when an EPC page needs to be added to the enclave (via EADD), the security monitor selects a page from the free list and constructs the corresponding page table entry (PTE).


In cases when the enclave's memory accesses result in a page fault, the CPU triggers a trap to the security monitor. The security monitor forwards the page fault information, including the address of the faulting page, to the \sysname-driver. If the \sysname-driver determines that a page frame needs to be allocated for the enclave, it requests the security monitor to allocate a physical page within the EPC. Additionally, the security monitor constructs the corresponding PTE in the enclave's page table. Once the page fault is successfully handled, the security monitor restores the enclave's execution context.
Currently, \sysname does not support swapping the enclave's pages to regular memory or disk storage.

\ignore{
\begin{figure}
    \centering
    \includegraphics[width=\linewidth]{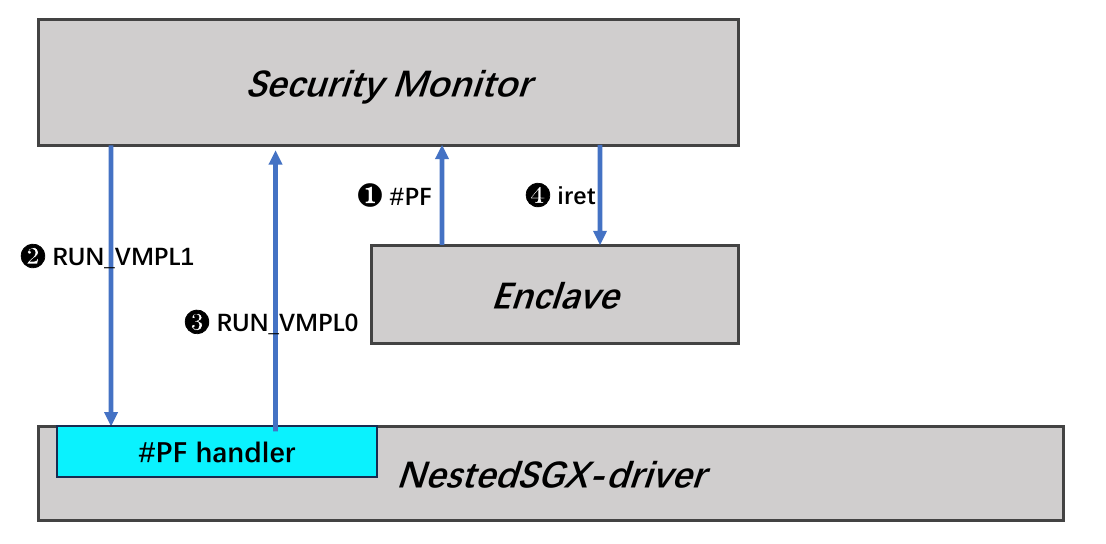}
    \caption{Handling Page faults in \sysname.}
    \label{fig:pagefault}
\end{figure}
}

\para{Isolation of enclaves}
It is also important to prevent an enclave from tampering with the security monitor or the guest OS. This ensures protection against potentially malicious enclave (i.e., enclave malware) or enclaves that may have vulnerabilities that could be exploited.
One approach to address this concern is to place the enclave at a distinct VMPL level, separate from VMPL0 or VMPL1, and enforce the appropriate VMPL permissions to contain the enclave code. However, the transition is mediated by the security monitor, necessitating at least two VMPL switches. As a result, transitioning between the application and the enclave becomes costly.

Instead, the enclave is positioned in user mode at VMPL0 and isolated using traditional page table-based isolation. Specifically, \sysname ensures that the enclave's page table does not map to any gPA belonging to the security monitor or the guest OS. As a result, transition between the application and the enclave only requires one VMPL switch.



\subsection{Measurement, attestation and sealing}
\label{subsec:attestation}

When the CVM image is loaded and measured, the page attributes, including the access permissions of all VMPLs, are included in the measurement. Any attempts to modify the code and data in VMPL0 and VMPL1, or to manipulate the access permissions of guest pages during CVM initialization, will be reflected in the measurement.
After the CVM is initialized, software operating at any VMPL level can initiate an attestation report request by sending a message to the AMD-SP firmware. The request structure comprises the VMPL level and 512 bits of space for user-provided data, which will be incorporated into the attestation report signed by the AMD hardware. The local/remote attestation and sealing mechanisms of \sysname closely mirror those employed by Intel SGX.
This is accomplished by emulating the corresponding instructions (i.e., EGETKEY and EREPORT) within the security monitor.

\begin{figure}
    \centering
    \includegraphics[width=\linewidth]{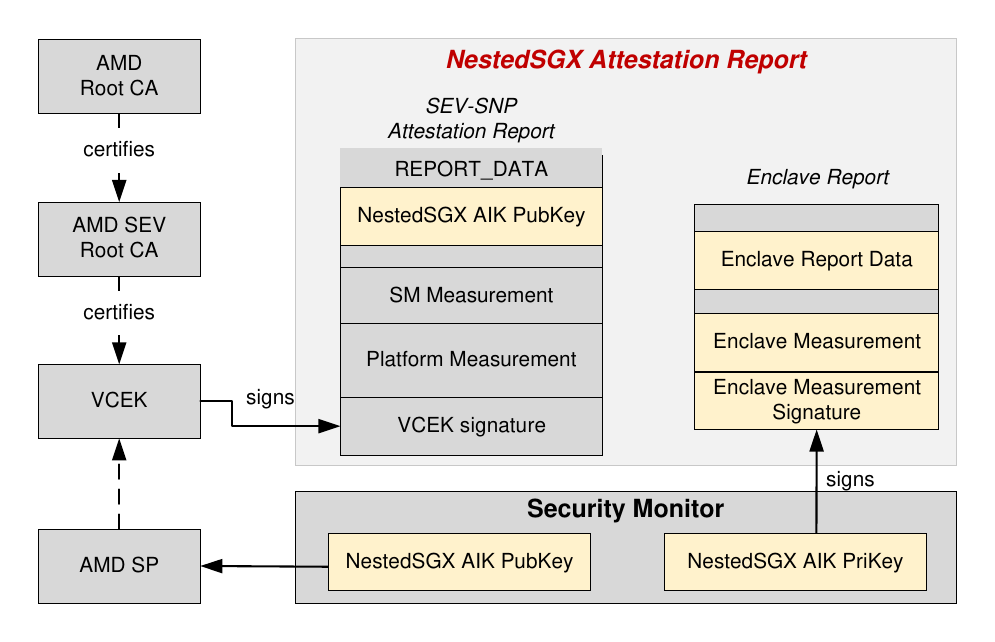}
    \caption{The \sysname attestation report consists of two parts: the SEV-SNP attestation report signed by VCEK, and the enclave report signed by \sysname AIK. The \sysname AIK is generated within the security monitor and subsequently bound to the SEV-SNP hardware by placing its public key as part of the SEV-SNP attestation report.}
    \label{fig:attestation}
\end{figure}

\para{Chain of trust}
When the CVM is first initialized, the security monitor randomly generates a key pair, which serves as the attestation identity key (AIK) of \sysname. The AIK is only required to be generated once and can be reused across different CVM boot cycles. To establish the binding between \sysname's AIK and the CVM, we include the digest of the public part of the attestation key in the user-data field of the attestation report request (Fig.~\ref{fig:attestation}), while the private key never leaves the memory of the security monitor.

The security monitor initiates the generation of an \textit{SEV-SNP attestation report} by dispatching a SNP\_REPORT\_REQ message to the AMD-SP hardware.
The request message between the security monitor and AMD SP is encrypted using the appropriate VM Platform Communication Key (VMPCK) for VMPL0. 
Upon receiving the encrypted message, the AMD-SP hardware decrypts it, verifies its integrity, and subsequently responds with an attestation report, which is signed with the SEV-SNP attestation key (Versioned Chip Endorsement Key, VCEK). The report includes the platform measurement, the security monitor measurement, the VMPL level, and the public key of \sysname's AIK.
The \sysname AIK is exclusively accessible to the quoting enclave (QE) and is subsequently used to sign the report of the enclave. This signed enclave report, along with the SEV-SNP attestation report, collectively constitutes the final attestation report in \sysname.

Upon receiving the final attestation report, the remote party verifies the certificate chain to ensure that the SEV-SNP attestation report is signed by VCEK. This process confirms that the report originates from an authentic AMD processor, has been executed within a CVM, and that the measurements of both the platform and the security monitor are accurate. Furthermore, he gains confidence that the AIK generated by \sysname is produced and retained within the security monitor. Finally, he verifies that the enclave report, which includes the enclave measurement, is signed with \sysname's AIK, concluding the attestation process.

\para{Sealing}
The security monitor can request keys from the AMD-SP by sending a MSG\_KEY\_REQ message to the AMD-SP. Upon receiving the request, the AMD-SP derives a key for the guest from a root key, which is responded to the security monitor and can be used as a sealing key. The sealing key can then be extended to support sealing of the enclave’s secret, binding the key to the enclave’s measurement. As part of the MSG\_KEY\_REQ message, the security monitor specifies that the derived key is used by VMPL0 and the AMD-SP mixes the VMPL information into the derived key.
Although the guest OS running at VMPL1 can also request keys with the MSG\_KEY\_REQ message, it is not allowed to specify VMPL0 in the message, because the hardware ensures that the specified VMPL is greater than or equal to the current VMPL. Therefore, the guest OS cannot derive the same key as the one derived by the security monitor.



\section{Implementations}
\label{sec:impl}


\begin{table}
\caption{\sysname software stack: the security monitor extends the SVSM framework and is mostly written in Rust and assembly. The \sysname-driver and Intel SGX SDK are mostly written in C/C++ and assembly.} 
\label{table:loc}
\centering
\begin{tabular}{C{2.1cm} C{4.0cm} C{1.6cm}} \toprule
    \textbf{Component} & \multicolumn{1}{c}{\textbf{Description}} & \textbf{Line of Code} \\ \midrule
    Security monitor & Emulation of SGX data structures, instructions and AEX& 5,500  \\ \midrule
 \sysname-driver & Handling switches between App and security monitor & 800 \\ \midrule
    SGX SDK \& Occlum & Replacing SGX instructions with IOCTL and system calls, HotCalls & 1,200 \\ \midrule
    \textbf{Total} &  & \textbf{7,500} \\ \bottomrule
\end{tabular}
\end{table}

Our implementation of \sysname is based on the software stack provided by AMD, which is open-sourced on GitHub~\cite{amdese}. The code consists of several components, including QEMU, Open Virtual Machine Firmware (OVMF), and Linux kernels for both host and guest environments. To implement \sysname, we extend the existing open-source Linux Secure VM Service Module (SVSM) framework~\cite{svsm2022spec,svsmgithub}, adding around 5,500 lines of code to the SVSM implementation. For VMPL switches, \sysname employs the standard Model Specific Register (MSR) protocol outlined in the Guest-Host Communication Block (GHCB) specification~\cite{ghcb}. The MSR protocol notifies the KVM to switch to the targeted VMPL, without any modification of the KVM.

\para{Security monitor}
We based our implementation on the main branch of SVSM without CPL3 support, and added our own support for running user-space enclaves.
The security monitor emulates SGX leaf functions based on requests from the \sysname-driver. To accommodate these requests, a new type of SVSM protocol called SGX protocol was introduced within the SVSM framework~\cite{svsm2022spec}. 
The security monitor effectively enters a request loop within the SVSM framework, allowing it to continue execution upon the arrival of subsequent requests. The security monitor handles a memory pool acting as the EPC, and faithfully emulates SGX data structures and operations in strict adherence to the SGX specification. 
We registered customized fault and exception handlers for the enclave, emulating the AEX event. Furthermore, we configured the system call handler by setting the LSTAR MSR to handle the EEXIT leaf function.

To enter the enclave, the security monitor performs a series of actions. It switches to the enclave's page table and stack, and then executes the \texttt{sysret} instruction with RCX pointing to the entry address, effectively emulating the behavior of EENTER and ERESUME leaf functions. The EEXIT leaf function, on the other hand, is replaced with the \texttt{syscall} instruction. In this case, the system call handler within the security monitor manages the VMPL switches. In order to handle AEX events, the security monitor directly accesses the enclave's State Saving Area (SSA) by disabling Supervisor Mode Access Prevention (SMAP). This allows the security monitor to handle AEX events efficiently.
Since the security monitor and the enclaves have separate page tables, we put the privilege switching code in a trampoline page that is mapped in both page tables, which is similar to the implementation of kernel page table isolation (KPTI)~\cite{gruss2017kaslr}. 

\para{\sysname-driver}
The \sysname-driver is responsible for managing the transitions between the App and the security monitor. When transitioning to the security monitor, it handles the SGX instruction emulation request and switches to VMPL0 using the SVSM protocol. The driver exposes the /dev/\sysname device to the App, which allows the App to invoke the request using the IOCTL interface, e.g., when entering the enclave with EENTER or ERESUME.
In cases where the execution is returned from the enclave to the application, such as after an AEX, OCALL request, or after processing the ECALL request, the driver uses the \texttt{sysret} instruction to resume the execution of the application.

\para{Intel SGX SDK}
We made modifications to the official Intel SGX SDK to replace the SGX instructions. Our implementation retains the same parameter semantics and orders as SGX for compatibility purposes. Specifically, instructions intended for use within the enclave, such as EEXIT, EREPORT and EGETKEY, were replaced with the \texttt{syscall} instruction, which allows for invocation of the security monitor. On the other hand, instructions meant for use within the App and guest OS, such as EENTER, ECREATE and EINIT were replaced with IOCTL calls, which trigger the appropriate functionality within the \sysname-driver.

With the aforementioned design, the majority of SGX programs can run on \sysname without requiring source code modifications. All the sample code projects provided by the official Intel SGX SDK, except SampleAEXNotify, ran without any issues on \sysname.
Furthermore, we have adapted the Rust SGX SDK~\cite{wang2019towards} and the Occlum library OS~\cite{shen2020occlum} to the \sysname platform.
Building upon Occlum, we implemented the HotCalls optimization~\cite{weisse2017regaining}, allowing certain OCALLs to be processed asynchronously within the enclave without exiting it.\footnote{We did not cover all OCALLs but focused on optimizing 15 frequently used ones, such as \path{occlum_ocall_sendmsg}, \path{occlum_ocall_recvmsg}, \path{occlum_ocall_clock_gettime}, \path{occlum_ocall_posix_memalign}, \path{occlum_ocall_free},  \path{u_sgxprotectedfs_fread_node} and \path{u_setsockopt_ocall} etc.}



\para{Intra-enclave isolation}
Since Intel SGX only supports the monolithic model, attempts have been made to support intra-enclave isolation, such as Nested Enclave~\cite{park2020nested} and LIGHTENCLAVE~\cite{gu2022hardware}. However, these solutions necessitate hardware modifications that go beyond Intel SGX's capabilities. We incorporated support for intra-enclave isolation, leveraging the hardware capabilities of memory protection keys for userspace (PKRU). We offer four PKRU APIs as syscalls for enclaves: \path{pkey_set}, \path{pkey_alloc}, \path{pkey_free}, and \path{pkey_mprotect}. Additionally, we provide two syscalls that allow enclaves to enable or disable PKRU mechanisms.





Initially, the enclaves invoke the \path{pkru_enable} function, informing the security monitor to establish the key manager. Subsequently, the enclaves can use the \path{pkey_alloc} and \path{pkey_set} functions to obtain a key from the manager and make use of it through \path{pkey_mprotect}.
%
On the \path{pkey_mprotect} function, the security monitor checks with the key manager to verify if the key has been assigned to the enclaves. If valid, the monitor updates the PKRU-bits of corresponding PTEs by traversing the enclaves' page tables. The key manager faithfully records and updates key information, including key attributes and the protected virtual address of enclaves. When the enclaves invoke the \path{pkru_free} function, the security monitor clears the PKRU-bits of the relevant PTEs. Lastly, the enclaves can call the \path{pkru_disable} function to instruct the security monitor to release the key manager.

\section{Security Analysis}
\label{sec:security}


We conduct the security analysis by enumerating the attack vectors under the threat model outlined in Section~\ref{subsec:threat}, and describing how \sysname defends against them.

\para{Untrusted App, \sysname-driver and guest OS}
The App and enclave have separated page tables, where the enclave's page table is managed by the security monitor and not accessible to the guest OS. This prevents various page table based attacks such as address mapping attacks (i.e., manipulating the enclave's address mappings) and side channels based on page fault or the access/dirty bits\footnote{Side channel attacks based on nested page tables (NPTs) are still possible, since the NPTs are managed by the untrusted hypervisor.}. 

The \sysname-driver is solely tasked with the switching of the VMPL for the emulation of SGX instructions.
It may attempt to modify the emulated instruction and its parameters, which however is equivalent to running a different SGX instruction. As the instruction is expected to be run by the untrusted App, this does not introduce additional attack surfaces. On the other hand, it may try to hijack the control flow after returning from the enclave, e.g., after EEXIT or AEX, however the control flow can be similarly manipulated by the malicious OS in SGX as well.

\para{Untrusted host VMM}
The VMPL switching requests are delivered to the host VMM via the MSR protocol. The host VMM is expected to switch to the targeted VMPL using the VMRUN instruction with the VMSA corresponding to the targeted VMPL as the parameter. In this process, the SEV-SNP hardware prevents it from tampering with the VMSA, which contains the state of the VM and the VMPL level. Upon request, the host VMM may try not to switch the VMPL or switch to a different VMPL, causing denial-of-service attacks, which is out of the scope of our paper.

During VMPL switches, the host VMM observes the patterns of ECALLs, OCALLs and AEXes. These leakages also exist in Intel SGX, if the App and the host OS are untrusted. Mitigating the side channels associated with enclave switches are also out of the paper's scope.

\para{Untrusted enclave} Although the enclave operates at the highest VMPL, it is not allowed to access the gPA belonging to lower VMPLs, such as the lower VMPL's VMSA. This is achieved by page table based isolation. Only the enclave’s pages are mapped in its page table which is being managed by the security monitor. The hardware also prevents the enclave from overwriting the VMPL permissions of guest pages, since the instruction to do so (RMPADJUST, Sec.~\ref{subsec:snp}) is privileged and can only be executed by the security monitor.

In any case, the enclave is not allowed to bypass the security monitor. For example, it cannot directly transmit data to the host without the intervention of the security monitor. Specifically, the instruction used for the MSR protocol (i.e., \texttt{wrmsr}) is a privileged instruction and can only be executed by the security monitor. Another venue of triggering VMEXIT by the enclave is through the unprivileged \texttt{vmgexit} instruction. However, the enclave cannot explicitly pass any data to the host since the security monitor does not allow shared memory between the enclave and the host. The side channel through intentionally triggering VMEXIT is out of the paper's scope.


For the enclave to enter the security monitor, the control flow is directed to a fixed location as specified by the LSTAR MSR of VMPL0. On faults and exceptions, the control flow is directed to the fault handler as specified in the Interrupt Descriptor Table (IDT). This prevents the enclave from arbitrarily diverting the control flow within the security monitor.

The enclave may try to mount DoS attacks on the guest OS, e.g., by a busy loop that never returns, since the guest OS is not allowed to inject an interrupt into VMPL0. \sysname prevents the attack by preempting the enclave execution using timer interrupts and triggering AEX events. We leave the design of proactively receiving interrupt request from the guest OS as future work.






\para{Chain of trust analysis}
The \sysname report includes the SEV-SNP attestation report and enclave report. Any attempt to tamper with the security monitor will result in a change to the SEV-SNP attestation report, thereby ensuring the security monitor is reliable and performs as expected. Specifically, the security monitor generates \sysname's AIK and maintains its private key inaccessible to other components. It associates \sysname's AIK with the platform by including the digest of the AIK's public key in the SEV-SNP attestation report.

While the enclave operates at VMPL0, it is prohibited from directly obtaining the SEV-SNP attestation report. 
Otherwise, the enclave may incorporate a counterfeit AIK into the report, which could compromise the integrity of the trust chain, particularly if the enclave deliberately exposes the private key of the AIK.
This protection is enforced by the fact that SNP\_REPORT\_REQ can only be initiated from the kernel mode, which is the security monitor. More specifically, SNP\_REPORT\_REQ necessitates shared memory with the host VMM for passing request and response messages, but we intentionally disable shared memory between the enclave and the host VMM.
The hardware also prevents the guest OS (operating at VMPL1) from obtaining an SEV-SNP attestation report for VMPL0. Specifically, the desired VMPL is provided by the guest in the \path{SNP_REPORT_REQ} message, and the guest can only generate attestation reports for VMPLs that are \textit{greater than or equal to the current VMPL}~\cite[Sec. 7.3]{sev-snp-firmware}.

The security monitor employs the SNP\_GUEST\_REQUEST command to solicit the SEV-SNP attestation report. This command establishes a trusted channel between the security monitor and the AMD-SP firmware, encrypted with AES-GCM authenticated encryption using the VMPCK associated with VMPL0. Each message contains a sequence number. Although the VMM mediates the communication between the CVM and the firmware, it is unable to modify or drop these messages without being detected, nor can it access the plaintext of the messages.





\section{Evaluations}
\label{sec:eval}


\subsection{Experimental setup}
We deployed \sysname on a server with two 
{AMD EPYC 7543} CPUs (two threads per core, total of 128 logical cores) with 64 GB DDR4 RAM. Every CVM was allocated 4 vCPUs and 4 GB RAM, and ran Ubuntu 22.04 with kernel version 6.5.0-snp-guest (svsm-preview-guest-v3 branch provided by AMD).
We configured 512 MB gPA for the security monitor and the enclaves.
The modified Linux SGX SDK was based on version 2.20, and the Occlum version was version 0.29.7. We used the main branch of the linux-svsm framework (commitID: 8d518f1). The SVSM framework and the security monitor were compiled with Rust in the default mode (-O0). All programs were compiled with GCC 11.4.0 and the same optimization level (-O2).

The baseline was conducted on the same AMD server running 
in the SGX SDK simulation mode (sim mode), which does not provide any security guarantees. Specifically, we used the same code base under the same optimization level and compiling options. Notably, as a result of the HotCalls optimization, applications running on Occlum experienced improved performance without the need for any modifications to their source code. We were not able to reproduce the experiment reported in vSGX, and the vSGX results are taken directly from the published paper~\cite{zhao2022vsgx}. {The Intel platform for comparison was equipped with an Intel Xeon Platinum 8369B CPU
and 64 GB RAM.}

\subsection{Micro-benchmarks}

\begin{table}[t]
    \centering
    \caption{Latencies of emulated SGX leaf instructions. The emulation of ENCLS leaf instructions require VMPL switches, which dominates the cost, while most ENCLU leaf instructions are emulated entirely at VMPL0. The slower performance of ENCLU leaf instructions on \sysname is attributed to the use of slower Rust implementations in the cryptographic crates.}
    \label{tab:leaf}
    \tabcolsep=0.45cm
    \begin{tabular}{c c c c}
        \toprule 
         &    &  \textbf{vSGX}\tnote{$\dagger$} &  \textbf{\sysname} \\
        \midrule
        \multirow{10}*{ENCLS} & ECREATE  &  3,719 us    & 8.4 us   \\
        ~ & EADD   &  1,421 us   &  8.0 us \\
        ~ & EEXTEND   &  987 us   &  33 us \\
        ~ & EINIT   &  811 us   &  46 us \\
        ~ & EREMOVE   &  1,014 us   &  7.8 us \\
        ~ & EAUG   &  990 us   &  7.9 us \\
        ~ & EBLOCK   &  841 us   &  9.2 us \\
        ~ & ELDB/ELDU   &  1,958 us   &  9.3 us \\
        ~ & EMODPR   &  1,071 us   &  9.9 us \\
        ~ & EWB   &  1,819 us   &  7.9 us \\
        \midrule
        \multirow{2}*{ENCLU} & EGETKEY  &   5.0 us   & 17 us   \\
        ~ & EREPORT   &  19 us   &  30 us \\
        \bottomrule
    \end{tabular}
    
\end{table}


\para{SGX leaf instructions}
We measured the emulated SGX leaf instructions in \sysname, averaging the results over 10,000 runs. As shown in Table~\ref{tab:leaf}, all ENCLS leaf instructions on \sysname are significantly faster than vSGX. This is attributed to the requirement of VMPL switches for emulating ENCLS leaf instructions on \sysname, whereas vSGX necessitates more costly cross-VM communications.
Both EGETKEY and EREPORT on \sysname are slower than those on vSGX, which we confirmed is due to the use of the Rust \texttt{rust-crypto} crate that is slower than the implementation used by vSGX. Notably, EGETKEY and EREPORT are not heavily invoked in practice.

\ignore{
\begin{table}[t]
    \small\centering
    \caption{Latencies of SGX primitives. A VMPL switch round-trip costs about 19,400 cycles on our platform, while the SGX SDK routine costs 4000--5500 cycles. The rest is the cost of our unoptimized implementation of the security monitor in emulating SGX instructions (about 9,000 cycles). The Intel platform was equipped with an Intel Xeon E3-1270 v6 CPU and 64 GB RAM.}
    \label{tab:primitive}
    \begin{tabular}{c C{2.2cm} c c}
        \toprule 
         &   Intel SGX hardware mode  &  Simulation (insecure) mode  &  \sysname\\
        \midrule
        ECALL  &  14,437    & 5,500 &  33,584  \\
        \midrule
        OCALL   &  12,586   &  3,927 & 32,014\\
        \bottomrule
    \end{tabular}
\end{table}
}

\begin{table}[t]
    \centering
    \caption{Latencies of context switches. The cost of \sysname includes the following parts: a VMPL switch round-trip which costs about 19,400 cycles on our platform, the SGX SDK routine which costs 4000--5500 cycles, and the rest is the cost of our unoptimized implementation of the security monitor in emulating SGX instructions (including an EENTER and an EEXIT, which cost about 9,000 cycles).}
    \label{tab:primitive}
    \tabcolsep=0.18cm
    \begin{tabular}{cccc}
        \toprule 
     &   \textbf{Intel SGX} &  \textbf{vSGX}\tnote{$\ddagger$}  &  \textbf{\sysname}\\
        \midrule
        ECALL  &  10,988 cycles (4.1 us)    & $\approx 1,500$ us &  33,584 cycles (12 us) \\
        \midrule
     OCALL   &  9,337 cycles (3.5 us)   & - & 32,014 cycles (11 us) \\
        \bottomrule
    \end{tabular}
\end{table}


\para{Context switches}
We measured the latency of ECALLs and OCALLs in \sysname. The test runs empty edge calls with no explicit parameters 100,000 times and takes the average value. As shown in Table~\ref{tab:primitive}, the latency of context switches in \sysname is $1.9\times$ -- $2.1\times$ higher than Intel SGX, but is still two
orders of magnitude faster than vSGX. Notably, existing works on Intel SGX, including EleOS~\cite{orenbach2017eleos}, HotCalls~\cite{weisse2017regaining} and Switchless calls~\cite{tian2018switchless} can be further applied to \sysname to reduce the cost brought by context switches. 

\ignore{
\begin{figure*}[t]
    \centering
    \begin{subfigure}{0.45\textwidth}
        \centering
        \includegraphics[width=\textwidth]{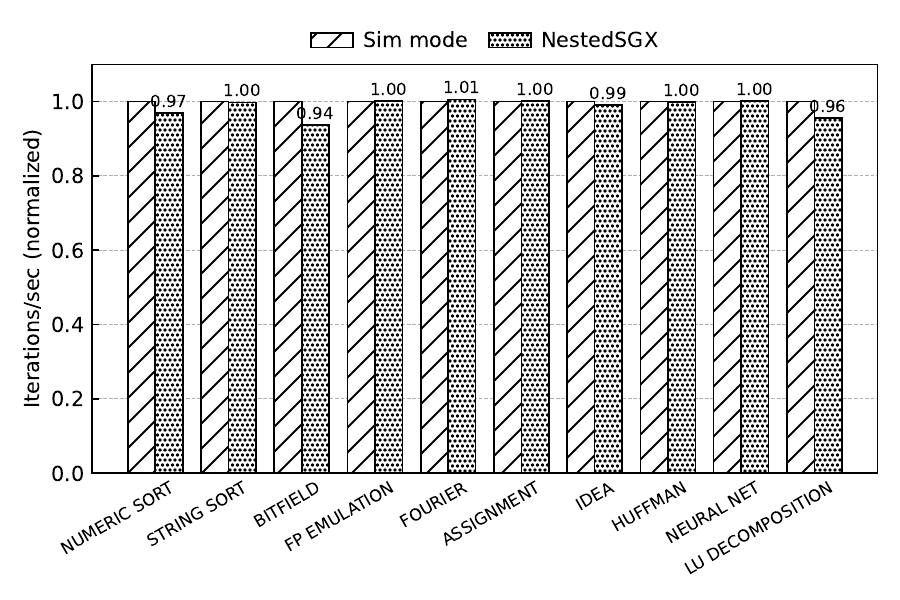}
        \caption{Linux/Unix nbench}
        \label{fig:nbench}
    \end{subfigure}
    \hspace{5pt}
    \begin{subfigure}{0.45\textwidth}
        \centering
        \includegraphics[width=\textwidth]{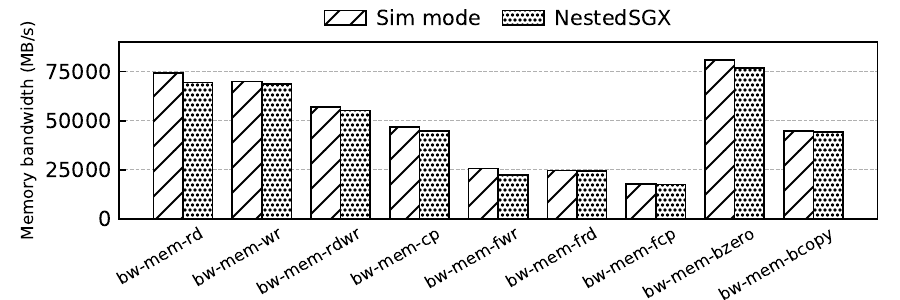}
        \caption{Lmbench}
        \label{fig:lmbench}
    \end{subfigure}
    \caption{Micro-benchmarks\wenhao{replace lmbench}.}
    \label{fig:micro-bench}
\end{figure*}
}

\begin{figure*}
    \centering
    \begin{subfigure}{0.45\textwidth}
        \centering
        \includegraphics[width=\textwidth]{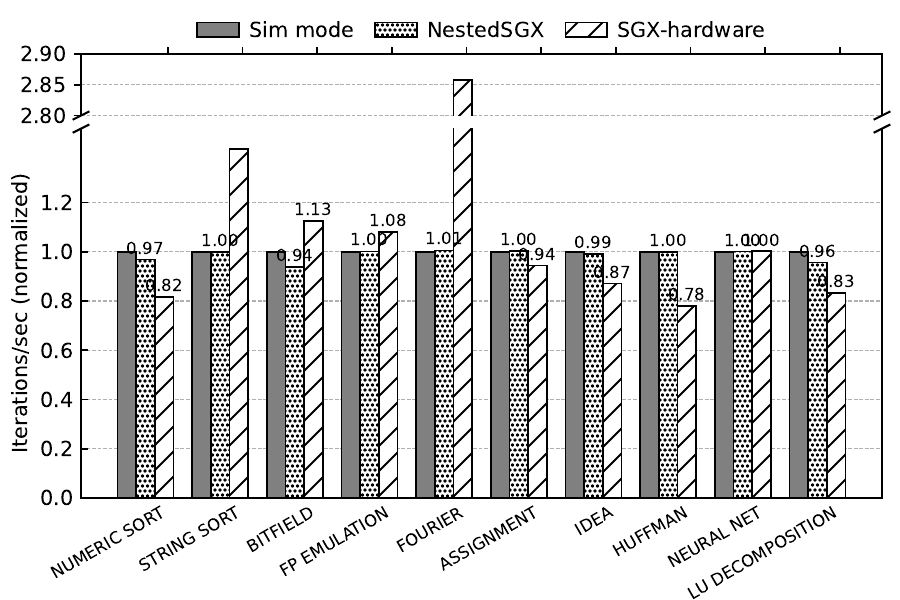}
        \caption{Linux/Unix nbench}
        \label{fig:nbench}
    \end{subfigure}
    \hspace{5pt}
    \begin{subfigure}{0.45\textwidth}
        \centering
        \includegraphics[width=\textwidth]{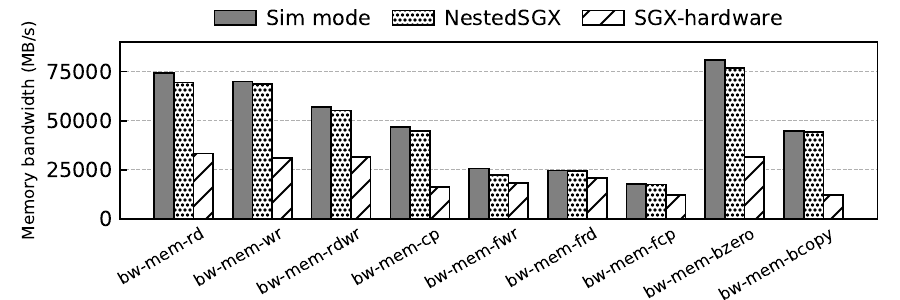}
        \caption{Lmbench}
        \label{fig:lmbench}
        \includegraphics[width=\textwidth]{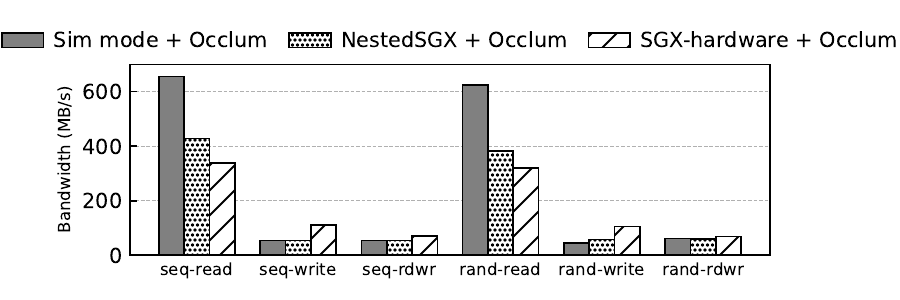}
        \caption{FIO}
        \label{fig:fio}
    \end{subfigure}
    \begin{subfigure}{\textwidth}
        \centering
        \includegraphics[width=\textwidth]{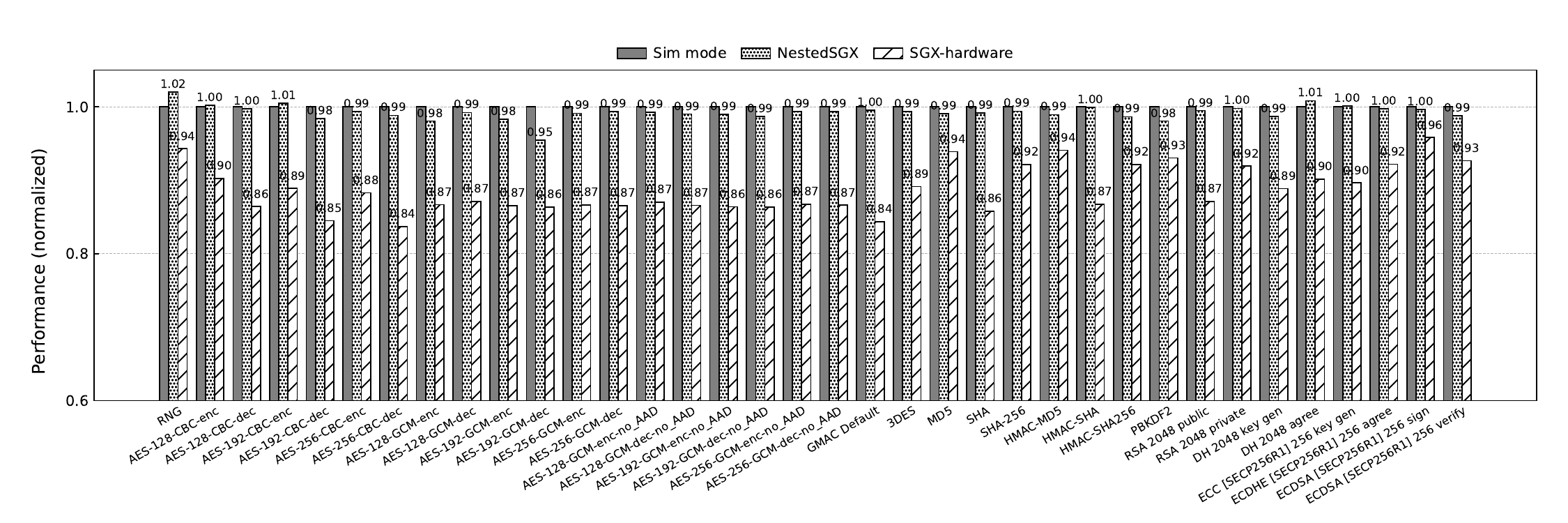}
        \caption{Wolfssl}
        \label{fig:wolf}
    \end{subfigure}
    \caption{Micro-benchmarks.}
    \label{fig:micro-bench}
\end{figure*}

\para{Linux/Unix nbench}
Linux/Unix nbench is a benchmark suite that focuses on evaluating the performance of a computer system's CPU, FPU, and memory system~\cite{nbench}. 
We utilized an adapted version of nbench, namely SGX-NBench~\cite{sgx-nbench} to evaluate the computation 
performance of \sysname. 
Fig.~\ref{fig:nbench} demonstrates that \sysname introduces an overhead of approximately 1.3\% on average over the baseline, 
which we believe mostly attributes to the context switches introduced by SGX-NBench during the evaluation.


\para{Lmbench}
Lmbench is a benchmark suite commonly employed to assess various aspects of system behavior. In our evaluation, we utilized the SGX version of Lmbench, known as SGX-bench~\cite{hasan2020port}, specifically employing the \texttt{bw\_mem} test to evaluate the memory bandwidth.
We conducted 10,000 repetitions of the test and recorded the average values. The memory block size for every access was set to 2 MB. The results, presented in Fig.~\ref{fig:lmbench}, indicate that the memory bandwidth achieved is 98.7\% of the baseline.



\para{WolfSSL~\cite{wolf}}
We performed an evaluation of WolfSSL with Intel SGX~\cite{wolfsslsgx}. This benchmark primarily focuses on computation-intensive tasks, such as encryption, decryption, digests, and signature verification.
As illustrated in Fig.~\ref{fig:wolf}, \sysname introduces an average overhead of about 0.78\% compared to the baseline.

\para{Flexible I/O Tester (FIO)~\cite{git:fio} with Occlum}
We conducted an evaluation of the I/O performance of \sysname by utilizing FIO (v3.28) with Occlum in the default configuration. Our focus was on benchmarking the bandwidth of randomized and sequential read and write operations to the disk, utilizing a single thread. We chose the direct I/O mode for sequential operations. Each test involved accessing a 256 MB file with a block size of 256 KB per access. We repeated the test 10 times, with each iteration lasting 100 seconds, and calculated the average bandwidth from all the runs.
As depicted in Fig.~\ref{fig:fio}, the I/O read performance of \sysname is approximately 65.3\% of the baseline, while the I/O write performance is similar to the baseline, possibly due to Occlum's optimizations to reduce the context switches in write operations.




\subsection{Real-world evaluations}


\begin{figure*}
\centering
    \begin{subfigure}{0.24\textwidth}
        \centering
        \includegraphics[width=\textwidth]{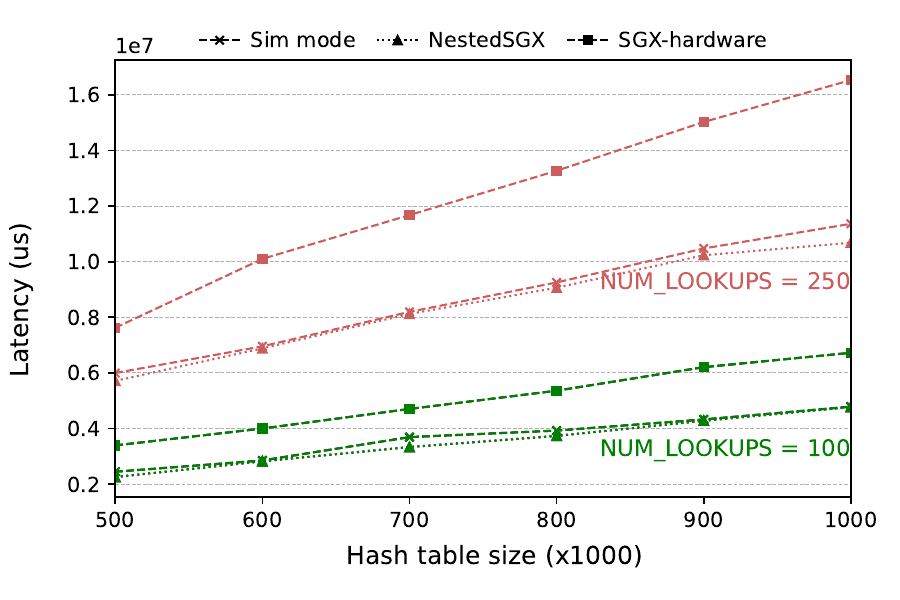}
        \caption{Hash join}
        \label{fig:hashjoin}
    \end{subfigure}
    \begin{subfigure}{0.24\textwidth}
        \centering
        \includegraphics[width=\textwidth]{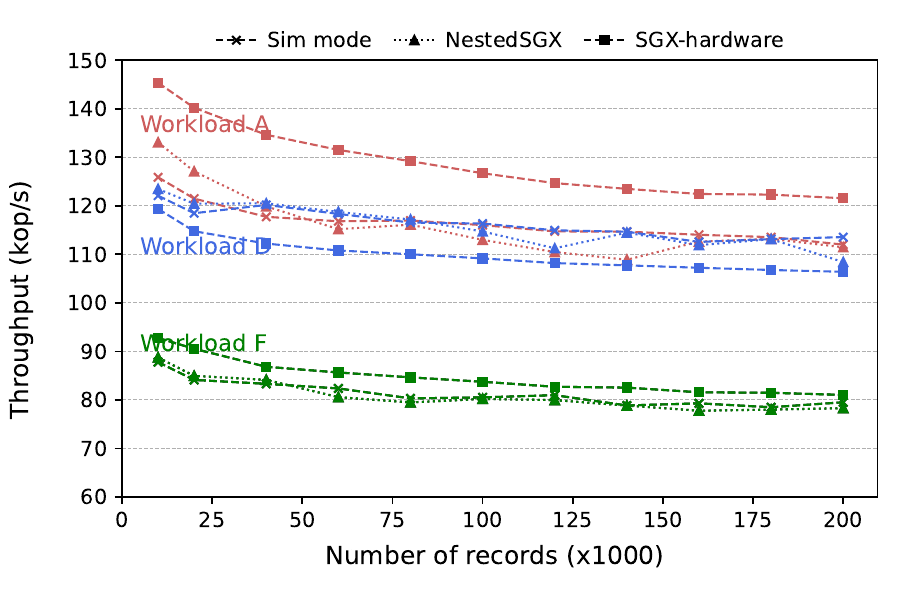}
        \caption{SQLite}
        \label{fig:sqlite}
    \end{subfigure}
    \begin{subfigure}{0.24\textwidth}
        \centering
        \includegraphics[width=\textwidth]{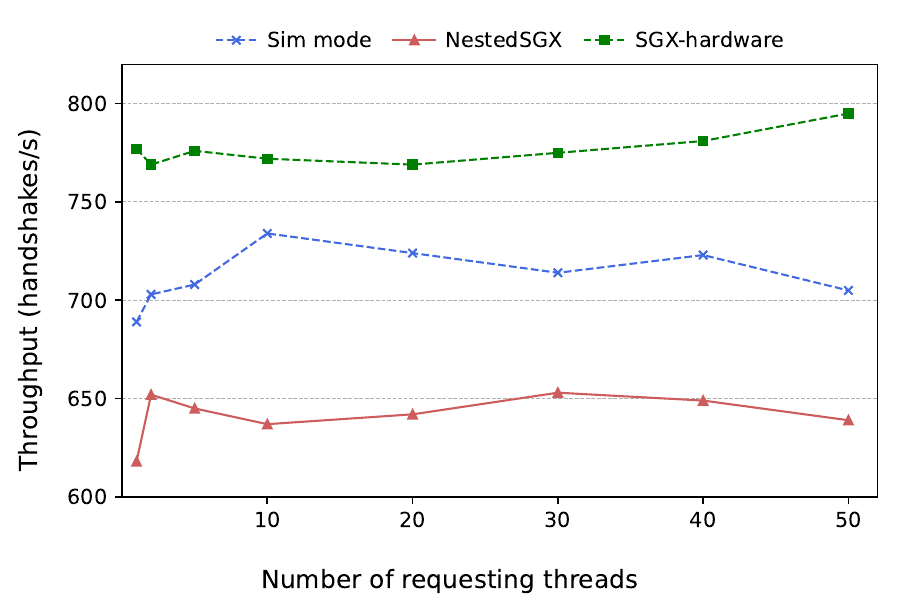}
        \caption{TLS server}
        \label{fig:tls}
    \end{subfigure}
    \begin{subfigure}{0.24\textwidth}
        \centering
    \includegraphics[width=\textwidth]{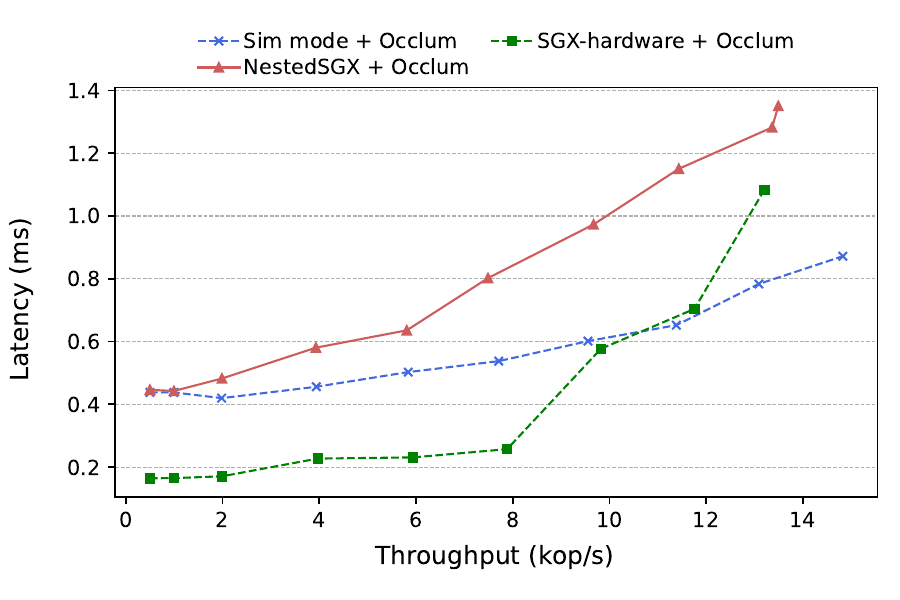}
        \caption{Redis with Occlum}
        \label{fig:redis}
    \end{subfigure}
    \caption{Real world application benchmarks.}
    \label{fig:macro-bench}
\end{figure*}



\para{Hash join}
Hash join is used to implement the ``equivalent-join'' operation in modern databases. It involves creating a hash table from rows of the first table and then probes it with rows of the second table. We used the open-source implementation of the algorithm in SGXGauge~\cite{kumar2022comprehensive,sgxgauga}. We varied the size of the first table (from 500K to 1M records) and fixed the second table (100K), effectively varying the memory and computation intensive nature of the workload. As shown in Fig.~\ref{fig:hashjoin}, the overhead of \sysname is negligible. 


\para{SQLite}
We utilized the publicly available version of SQLite (v3.19.3) operating on Intel SGX~\cite{sqlite}. 
To evaluate the memory performance of \sysname, we configured the database to function in-memory mode and incorporated the client within the enclave. We conducted performance evaluations across various record sizes, measuring the time required for 100,000 database operations. Specifically, we employed three representative workloads within the Yahoo! Cloud Serving Benchmark (YCSB) suite~\cite{ycsbworkload}: workload $A$ (update heavy with 50\% reads and 50\% updates), $D$ (read latest, i.e., delete old ones, insert new ones and read mostly the new ones), and $F$ (short range scan) for the evaluation. As shown in Fig.~\ref{fig:sqlite}, \sysname introduces about 
0.77\% overhead on average over the baseline.

\ignore{
\para{Nginx}
In this evaluation,

\para{Lighttpd}
the throughput is only 27.24\% of native, and 101.98\% of sim (seems not correct)

\para{Redis}
the throughput is only 90.71\% of native, and 98.69\% of sim (seems not correct)

\para{SGX-TaLoS}
Nginx web server running as App, and the TLS runs as the enclave; wrk2 utility for benchmarking

https://github.com/lsds/TaLoS
}
\para{TLS server}
We utilized the open-source SGX-OpenSSL project and a sample implementation of a TLS server in an enclave~\cite{sgxopenssl} for our evaluation. 
The TLS client operated within the native CVM and sent requests to the server over the local loop-back. To measure the average latency, we employed the tls-perf benchmarking tool~\cite{tlsperf} and set the number of requesting threads to 1, 2, 5, 10, 20, 30, 40, 50 respectively.
As illustrated in Fig.~\ref{fig:tls}, \sysname incurs approximately a 9.75\% overhead over the baseline.

\ignore{
\para{Lighttpd}
We deployed a Lighttpd server (v1.4.40) on \sysname with Occlum and employed the Apache HTTP benchmarking tool for the test. The benchmark involved running 100 concurrent clients over the local loopback to fetch web pages of various sizes (from 1 KB to 200 KB), allowing us to evaluate the throughput.
As depicted in Fig.~\ref{fig:lighttpd}, we observed a significant drop in throughput, reaching approximately 22\% of the baseline. This performance degradation was primarily caused by the frequent enclave mode switches.
}

\para{Redis}
We ran a Redis (v6.0.9) database server with
Occlum on \sysname. 
The database was configured as an in-memory setup, and we applied the YCSB workload A for our evaluation. We first loaded 5,000 records (in total 5 MB data), and subsequently executed 10,000 operations from 20 clients over the local loopback. We gradually increased the request frequency and measured the corresponding latency at different throughput levels.
As shown in Fig.~\ref{fig:redis}, \sysname achieved a throughput of 84.32\% (15.68\% overhead) compared to the baseline. The latency overhead was found to be 6.4\% under low load conditions and increased to approximately 55\% as the system approached its maximum throughput.

\ignore{
\begin{table}[t]
    \small\centering
    \begin{threeparttable}[b]
    \caption{Wolfssl’s wolfcrpyt benchmarks on vSGX and NestedSGX}
    \label{tab:leaf}
    \tabcolsep=0.45cm
    \begin{tabular}{c c c c}
        \toprule 
         &    \textbf{vSGX}\tnote{$\dagger$} &  \textbf{\sysname} & \textbf{Ratio} \\
         & MB/s & MB/s & \\
        \midrule
        RNG & 82.57 & 104.71 & 1.27 \\
        AES-128-CBC-enc & 187.36 & 313.48 & 1.67 \\
        AES-128-CBC-dec & 172.59 & 368.19 & 2.13 \\
        AES-192-CBC-enc & 156.95 & 271.18 & 1.73 \\
        AES-192-CBC-dec & 184.4 & 304.72 & 1.65 \\
        AES-256-CBC-enc & 139.01 & 236.01 & 1.70 \\ 
        AES-256-CBC-dec & 123.05 & 263.71 & 2.14 \\ 
        AES-128-GCM-enc & 54.10 & 70.41 & 1.30 \\
        AES-128-GCM-dec & 56.02 & 70.86 & 1.26 \\
        AES-192-GCM-enc & 54.36 & 68.03 & 1.25 \\ 
        AES-192-GCM-dec & 54.49 & 66.22 & 1.22 \\
        AES-256-GCM-enc & 51.78 & 66.14 & 1.28 \\ 
        AES-256-GCM-dec & 49.74 & 66.36 & 1.33 \\
        3DES & 22.60 & 32.88 & 1.45 \\
        MD5 & 296.77 &  686.11 & 2.31 \\
        SHA & 223.09 & 740.99 & 3.32 \\
        SHA-256 & 115.56 & 251.01 & 2.17 \\
        HMAC-MD5 & 377.70 & 683.24 & 1.81 \\
        HMAC-SHA & 381.57 & 738.22 & 1.93 \\
        HMAC-SHA256 & 164.82 &  249.09 & 1.51 \\ 
        \midrule
         & KB/s & KB/s & \\
        \midrule
        PBKDF2 & 9.49 & 30.19 & 3.18 \\
        \midrule
         & op/s & op/s & \\
        \midrule
        RSA 2048 public & 10264.09 & 8165.22 & 0.80 \\
        RSA 2048 private & 188.40 & 146.56 & 0.78 \\
        DH 2048 key gen & 378.24 & 348.72 & 0.92 \\
        DH 2048 agree & 614.50 & 351.941 & 0.57 \\
        ECC 256 key gen & 453.50 & 11365.96 & 25.06 \\
        ECDHE 256 agree & 1461.67 & 4148.12 & 2.84 \\
        ECDSA 256 sign & 3611.59 & 7447.38 & 2.06 \\
        ECDSA 256 verify & 1336.96 & 3746.32 & 2.80 \\

        \bottomrule
    \end{tabular}
    \begin{tablenotes}
    \item[$\dagger$] The vSGX results are taken from the published paper~\cite{zhao2022vsgx}.
    \end{tablenotes}
    \end{threeparttable}
\end{table}
}



\section{Limitations \& Future Works}
\label{sec:discuss}


\para{Limitations}
{Since our implementation is still in the prototype stage, \sysname has some limitations to fully support the SGX model. Firstly, the security monitor manages the enclave’s page tables, which prevents the guest OS from manipulating the enclave’s page tables for potential address mapping attacks. Since the App and enclave have separate page tables, the enclave cannot access the App’s memory, as their memory views may differ if the App’s page table is updated without synchronizing with the enclave’s page table. Therefore, features such as \path{user_check} and memory sharing are not currently supported by \sysname. Secondly, the \sysname-driver does not currently support swapping the enclave’s memory to disks.
Thirdly, the guest OS lacks the capability to inject an interrupt into VMPL0. This limitation imposes constraints on the OS's scheduling of enclave threads.
Finally, compatibility is currently provided on top of SGX SDK and Occlum. Thus, applications that are purely written in the SGX instruction set without any SDK or LibOS are not yet supported. We do not think that any of these limitations are inherent to \sysname, e.g., memory sharing could be supported if updates to the App’s page table were synchronized with the enclave’s page table via VMPL switches.}

\para{Supporting other VM-based TEEs}
Recently, ARM CCA introduces the support of different planes within a CVM, in which each plane is essentially a separated VM but they all share the same guest physical address space. Plane 0 is more privileged and can run a paravisor which manages the switches between other planes and can restrict the memory accessible by other planes~\cite{armplanes}. According to ARM's roadmap, the paravisor is designed to support secure services such as vTPM emulation. Similarly, we can re-purpose the paravisor to support running userspace enclaves.
TDX supports a similar feature named TD partitioning. Therefore, it is promising that \sysname can be extended to other CVM platforms. However, it is worth mentioning that the TDX module and CCA's Realm Management Monitor (RMM) are critical for the memory isolation and secure management of CVMs, and the customization of these components may not be allowed by chip companies. In fact, only the TDX module signed by Intel can run in the Secure Arbitration Mode (SEAM).


\para{Supporting other TEE abstractions}
\sysname adopts the SGX model as it is arguably the most prevalent TEE with process abstraction. The \sysname platform can be enhanced to accommodate other TEE abstractions, such as ARM TrustZone. Specifically, VMPL0 could run the trusted OS (e.g., OP-TEE) and trusted applications (TAs), providing secure services to the guest OS running at VMPL1, which is analogous to the normal world.
The SMC instruction is used to handle transitions between the secure world and normal world. In TrustZone, the SMC handler is within the trusted firmware. In our system, the SMC instruction needs to be replaced with a VMPL switching request, which transfers the execution to the trusted OS. This allows porting existing applications built on ARM TrustZone to SEV-SNP CVMs.
\section{Related Works}
\label{sec:related}

\para{Integrity Measurement Architecture (IMA)}
Integrity plays a vital role in ensuring system security by guaranteeing the exclusive loading of authentic software onto a machine. Measured boot, utilizing the Trusted Platform Module (TPM), securely captures measurements of all boot software, culminating in the loading and execution of the OS kernel. The Linux Integrity Measurement Architecture (IMA) expands upon the concept of measured boot by comprehensively recording all software executions and file accesses within the OS, securely storing them in the TPM~\cite{sailer2004design}. The Container-IMA (C-IMA) extends IMA to containers, enabling the measurement of container images and the runtime integrity measurement of container processes~\cite{luo2019container}.

In lieu of TPM, TZ-IMA extends the storage of measurements in TrustZone~\cite{song2022tz}. Recently, Intel introduced a solution of IMA in confidential cloud environments, with a focus on Intel TDX, enabling IMA of programs and applications running in Intel trusted domain (TD) during runtime~\cite{tdima}. 

However, IMA in general assumes the trust of the boot components, which includes the BIOS, bootloader and guest OS, leading to potentially large TCB and expanded attack surface. For example, it has been shown that IMA can be bypassed with a malicious block device~\cite{bohling2020subverting}. In contrast, the core root of trust for measurement of \sysname (i.e., the security monitor) operates at VMPL0 and we do not place trust on the guest OS. Moreover, the security monitor not only measures the integrity of both enclave code and data but also ensures that the enclave runs in an isolated environment from the guest OS, which is beyond the design goal of IMA.

\para{Establishing the enclave abstraction within TEEs}
With the emergence of Intel SGX, endeavors have been undertaken to enhance the interoperability between Intel SGX and other TEE platforms. Notably, Komodo~\cite{ferraiuolo2017komodo}, Sanctuary~\cite{brasser2019sanctuary}, and SecTEE~\cite{zhao2019sectee} offer support for enclave abstractions on ARM TrustZone. HyperEnclave~\cite{jia2022hyperenclave} re-introduces Intel SGX on standard AMD hardware through the use of a small and trusted hypervisor. In contrast to these works, \sysname sets itself apart by adopting a distinctive model. While the guest OS remains shielded from the untrusted hypervisor, it is not entirely trusted. Consequently, \sysname enables enhanced protection for enclaves within the feature-rich guest OS.
Most related to our work are vSGX~\cite{zhao2022vsgx} and Veil~\cite{ahmad2023veil}. A more detailed comparison with them is provided in Sec.~\ref{subsec:overview}.



\para{Security and performance enhancement using the VMPL feature}
Hecate~\cite{ge2022hecate} utilizes VMPL0 as an L1 hypervisor, operating within the CVM on top of the untrusted L0-hypervisor (host VMM). Hecate enables the migration of on-premises workloads by maintaining compatibility with unmodified VM images and implementing security policies, such as network firewalls.
Honeycomb~\cite{mai2023honeycomb} executes a validator within VMPL0, safeguarded from other software. The validator analyzes the binary code of a GPU kernel to validate that each memory instruction within the GPU kernel can only access specific memory regions,
employing static analysis techniques.
SVSM-vTPM~\cite{narayanan2023remote} employs VMPL to isolate the virtual TPM (vTPM) from the guest OS, ensuring the integrity of vTPM's functionalities and the security of the root key for remote attestation. However, all these solutions do not provide isolation for user programs from the untrusted guest OS.


\section{Conclusions}
\label{sec:conclusion}
Although confidential computing aims to protect the confidentiality and integrity of code and data within TEEs, it remains challenging to control code within CVMs since the feature-rich guest OS is free to load any code. This paper presents \sysname, a secure and efficient solution for maintaining control over the integrity of code and data in TEEs. \sysname creates hardware enclaves inside CVMs with VMPL, which ensures remote attestation of trusted and measured code. \sysname is compatible with Intel SGX and easily integrated into existing systems. 
\sysname incurs a small overhead in most real-world applications.

\section*{Acknowledgments}  
We would like to express our sincere gratitude to the anonymous reviewers and our shepherd. The project was supported in part by the National Key R\&D Program of China (Grant No. 2020YFB1805402), the National Natural Science Foundation of China (Grant No.62272452) and the research grant from Ant Group. Corresponding authors: Shoumeng Yan (\href{mailto:shoumeng.ysm@antgroup.com}{shoumeng.ysm@antgroup.com}) and Rui Hou (\href{mailto:hourui@iie.ac.cn}{hourui@iie.ac.cn}).

\balance
\bibliographystyle{IEEEtranS}
\bibliography{ref}

\section*{Artifact Appendix}

\begin{table*}[b]
\small\centering
    \caption{Summary of the benchmarks included in the artifact. We do not currently offer automatic scripts to plot the evaluations of FIO, Redis, and the TLS server as these benchmarks necessitate some manual effort during evaluation. However, you can use our provided scripts to plot versions of FIO and Redis that do not include HotCalls.}
    \label{tab:ae_cmp}
    \begin{tabular}{C{2.5cm} C{2.6cm} C{2.1cm} C{9cm}}
        \toprule    
        Experiments   & Figure/Table    & Estimated time &  Description     \\
    \midrule
    Leaf instructions & Table~\ref{tab:leaf}  &  1m  & \makecell[c]{The latency of emulated SGX leaf instructions.}  \\
        \midrule
    edge-calls & Table~\ref{tab:primitive}  &  10s  & \makecell[c]{The latency of ECALLs/OCALLs.}  \\
    \midrule
    NBench & Figure~\ref{fig:nbench} & 10m  & Performance scores of NBench inside the enclaves. \\
    \midrule
    Lmbench & Figure~\ref{fig:lmbench} & 10m  & Lmbench memory bandwidth inside the enclaves. \\
    \midrule
    FIO & Figure~\ref{fig:fio} & 10m  & Bandwidth of FIO read and write operations inside the enclaves. \\
    \midrule
    Wolfssl & Figure~\ref{fig:wolf} & 10m  & Throughput of Wolfssl algorithms inside the enclaves. \\
    \midrule
    Hash join & Figure~\ref{fig:hashjoin}  & 15m & \makecell[c]{Latency of hash join inside the enclaves.} \\
    \midrule
    SQLite & Figure~\ref{fig:sqlite}  & 15m & \makecell[c]{Throughput of in-memory SQLite database with different \\ number of records, under YCSB A workload.} \\
    \midrule
    TLS server & Figure~\ref{fig:tls}  & 15m & \makecell[c]{Throughput of TLS server inside the enclaves.} \\
    \midrule
    Redis  & Figure~\ref{fig:redis} & 20m  & \makecell[c]{Latency-throughput curve of Redis in-memory database server \\ inside Occlum LibOS with increasing request frequencies.} \\
        \bottomrule
    \end{tabular}
    
\end{table*}

\subsection{Abstract}

\sysname can support existing SGX toolchains (the adapted SGX SDK and Occlum library OS) and run SGX applications atop AMD SEV-SNP confidential VMs (CVMs). This artifact contains the binaries of the security monitor, and documentations on how to setup the \sysname environment. 

The artifact includes benchmarks for edge calls (i.e., ECALLs and OCALLs), and benchmarks for the real-world workloads, including NBench, SQLite and Redis etc. We offer scripts to replicate the results outlined in the paper, as summarized in Table~\ref{tab:ae_cmp}.


\ignore{\subsection{Contents}

\begin{packeditemize}
\item{\texttt{README.md}} describes the artifact and provides a road map for evaluation.
 \item{\texttt{host/}} contains RustMonitor binary, the Linux kernel module binary, and the scripts to install and enable HyperEnclave.
 \item{\texttt{server/}} contains the source code (or patches) and scripts of all experiments to run within the enclaves. We also provide a docker container with all dependencies installed.
 \item{\texttt{client/}} contains the benchmark scripts for network-based experiments (Lighttpd and Redis) to run on the client side. We also provide a docker container with all dependencies installed.
 \item{\texttt{plots/}} contains plotting scripts to generate figures from the experiment results.
 \item{\texttt{paper-results/}} contains the results shown in the paper.
\end{packeditemize}
}

\subsection{Hosting}

Check out \href{https://github.com/NestedSGX/nestedsgx-ndss25-ae/}{https://github.com/NestedSGX/nestedsgx-ndss25-ae/}.

\subsection{Requirements}

\para{Hardware requirements}:
\begin{packeditemize}
    \item 3rd Gen AMD EPYC processors with SEV-SNP enabled in the BIOS. 
    \item RAM $\ge$ 16 GB.
    \item Free disk space $\ge$ 200 GB.
\end{packeditemize}

\para{Software requirements}
\begin{packeditemize}
    \item Linux with the specified kernel version (i.e., \texttt{6.2.0-26-generic}) to build the NestedSGX environment. We recommend Ubuntu 22.04 which uses this version of kernel as the default. 
    \item Git.
    \item GCC 11.4.0.
    \item Rust tool chain nightly 1.71.
\end{packeditemize}

\subsection{Steps}
The Major steps to set up the environment and reproduce the results are listed below. For more detailed instructions, please refer to the guidelines outlined in the \texttt{README.md} file. 

\begin{packeditemize}
\item \para{Configure the environment}
Begin by setting up the environment following AMD's Linux SVSM project~\cite{svsmgithub}. This process involves building specific kernels for both the host OS and the guest OS, along with the corresponding QEMU and OVMF files..

\item \para{Replace the security monitor}
Substitute the original \path{svsm.bin} with our customized version to serve as the security monitor for initiating \sysname.

\item \para{Install the guest module}
Install our guest module on the guest OS. This kernel module facilitates redirecting control flow within \sysname.

\item \para{Prepare the Linux SGX SDK environment}
Set up the Linux SGX SDK environment to enable simulation mode in the guest VM. Once completed, you can run micro-benchmarks like NBench, Lmbench, Wolfssl, as well as macro-benchmarks including hash join, SQLite, and TLS server.

\item \para{Set up the Occlum library OS for FIO and Redis} Compile and install the modified Occlum to support FIO and Redis. Enhance the performance of these benchmarks by utilizing the HotCalls versions of both the Intel SGX SDK and Occlum.

\end{packeditemize}



\end{document}